\documentclass[aps, prd, nofootinbib, preprintnumbers, showkeys, preprint]{revtex4}
\usepackage[usenames,dvipsnames]{pstricks}
\usepackage{epsfig}
\usepackage{pst-grad} 
\usepackage{mathtools}
\usepackage{array}

\usepackage{amsmath}
\usepackage{graphicx}
\usepackage{longtable}
\usepackage{multirow}

\usepackage{amssymb}
\usepackage{makeidx}
\usepackage{gensymb}
\usepackage{amsfonts}
\usepackage{color}
\newcommand{\bal}{\begin{align}}
\newcommand{\eal}{\end{align}}
\newcommand{\lbr}{\left[}
\newcommand{\rbr}{\right]}

\usepackage{allpurpose}

\begin{document}

\title{Time delay of timelike particles in gravitational lensing of Schwarzschild spacetime}

\author{Junji Jia}
\email{junjijia@whu.edu.cn}
\affiliation{Center for Astrophysics \& Center for Theoretical Physics \& MOE Key Laboratory of Artificial Micro- and Nano-structures, School of Physics and Technology, Wuhan University, Wuhan, 430072, China}

\author{Haotian Liu}
\email{htliu@whu.edu.cn}
\affiliation{School of Physics and Technology, Wuhan University, Wuhan, 430072, China}


\begin{abstract}

Time delay in Schwarzschild spacetime for null and timelike signals with arbitrary velocity $v$ is studied. The total travel time $t_\mathrm{if}$ is evaluated both exactly and approximately in the weak field limit, with the result given as functions of signal velocity, source-lens and lens-observer distances, angular position of the source and lens mass. Two time delays, $\Delta t_v$ between signals with different velocities but coming from same side of the lens and $\Delta t_\mathrm{p}$  between signals from different sides of the lens, as well as the difference $\Delta t_{\mathrm{p}v}$ between two $\Delta t_\mathrm{p}$'s are calculated. These time delays are applied to the gravitational-lensed supernova neutrinos and gravitational waves (GW). It is shown that the
$\Delta t_v$ between different mass eigenstates of supernova neutrinos can be related to the mass square difference of these eigenstates and therefore could potentially be used to discriminate neutrino mass orderings, while the difference $\Delta t_{\mathrm{p}v}$ between neutrino and optical signals can be correlated with the absolute mass of neutrinos. The formula for time delay in a general lens mass profile is derived and the result is applied to the singular isothermal sphere case. For GWs, it is found that the difference $\Delta t_{\mathrm{p}v}$ between GW and GRB can only reach $1.45\times 10^{-5}$ second for very large source distance ($2\times 10^4$ [Mpc]) and source angle (10 [as]) if $v_{GM}=(1-3\times 10^{-15})c$. This time difference is at least three order smaller than uncertainties in time measurement of the recently observed GW/GRB signals and thus calls for improvement  if $\Delta t_{\mathrm{p}v}$ is to be used to further constrain the GW velocity.

\end{abstract}



\maketitle

\section{Introduction}

With the discovery of supernova neutrino (SNN) from SN1987A \cite{Hirata:1987hu, Bionta:1987qt}, the recent observation of gravitational wave (GW) signal \cite{Abbott:2016blz,Abbott:2016nmj,Abbott:2017oio,TheLIGOScientific:2017qsa} and more recent confirmation of neutrino emission from blazer TXS 0506+056 \cite{IceCube:2018cha,IceCube:2018dnn}, astronomy has certainly entered the multi-messenger era. In particular, the observation of the binary neutron star merger GW170817 and GRB 170817A \cite{TheLIGOScientific:2017qsa,GBM:2017lvd,Monitor:2017mdv} is a simultaneous observation of the Gamma-ray burst (GRB) and GW signals. The 1.74 [s] time difference between the GRB and GW signals constrains the speed difference of gravity and light, the Equivalence Principle and the physical properties of the central engine of the GRB \cite{GBM:2017lvd}. It can also put stringent constraints on the parameter space of general scalar-tensor \cite{Sakstein:2017xjx},
vector-tensor gravity \cite{Baker:2017hug}, dark energy \cite{Creminelli:2017sry,Ezquiaga:2017ekz} and dark matter models \cite{Boran:2017rdn}.
This time difference however, can originate from one or more of the three sources: (i) the generation region of the GW-GRB, (ii) the propagation path from source to local galaxy, and (iii) near the Galaxy until received by detectors. For signals from sources of high redshift such as GW170817, it is generally expected that the gravitational potential along propagation is more important than the potential at the originating or receiving sites.

If there exist large mass along the propagation path, then the signal will experience gravitational lensing (GL), regardless whether the signal is neutrino, GRB or GW, massive or massless. Many observables of the GL such as the apparent angles, the time delay between different images and the magnification can be used to deduce properties of the signal source, the signal itself and spacetime it went through \cite{Walsh:1979nx,Aubourg:1993wb,Alcock:2000ph,Gaudi:2008zq,Gould:2010bk,Oguri:2010ns,Treu:2010uj}.
The time delay between different images of the same kind of signal has a special advantage over the time delay between signals of different particles in the same event: the former do not suffer the uncertainty of emission time since these different images are from the same emission. Even if the emission times of this signal is not known exactly, the time delay produced during traveling are still valuable to deduce properties of the sources, the signal particle/wave or the  spacetime transmitted.

However, in the computation of the time delay of timelike particles, the formulas for null ones are usually used \cite{barrow1987lensing, Eiroa:2008ks, Fan:2016swi,Wei:2017emo,Yang:2018bdf}, i.e., the timelike nature of the massive particle/wave was not fully accounted. Although the neutrinos from supernova and GWs from mergers are usually relativistic, the time delay itself indeed is the difference between the total travel times, which are large quantities too. Therefore for high accuracy calculations, especially when the timelike particle is not that relativistic or the lensing is strong, the timelike nature of the particles shall be fully addressed.

To fulfill this purpose, in this paper we propose to study the time delays between signals with different velocities and between different images of same or different kind of signals, especially those with nonzero masses. For simplicity, we will concentrate on the Schwarzschild spacetime. Previously, the time delay of light has been extensively studied using different approaches in the weak and strong field limits in various spacetimes or gravities \cite{Richter:1982zz,Edery:1997hu,Bozza:2003cp,Sereno:2004,Jacob:2008bw,Bailey:2009me,Eiroa:2013,Sahu:2013,Wang:2014yya,Zhao:2016kft, He:2016cya, Zhao:2017cwk,Zhao:2017jmv,Deng:2017umx}. For massive particles, the time delay of massive photon was investigated in Ref. \cite{Glicenstein:2017lrm} to constrain photon mass. Ref. \cite{Baker:2016reh} studied the difference between time delay of massive neutrino and that of massless GW to place bounds on total neutrino mass and some cosmological parameters. We emphasize that our work is different from theirs in a few ways. First, unlike Ref. \cite{Baker:2016reh, Glicenstein:2017lrm} which works only in the relativistic limit of the signal and weak field limit of the lens, we computed the {\it exact} total travel time that works for arbitrary velocity, i.e., velocities not very close to $c$.
Secondly, even in the relativistic limit, the total time and time delay in these works are only to the order $\calco(c-v)^1$ while our approximation formulas can works to higher orders. Thirdly, we have computed two time delays, $\Delta t_v$ due to velocity difference and $\Delta t_\mathrm{p}$ due to path difference, instead of only $\Delta t_\mathrm{p}$. These time delays and their differences are used to constrain the mass ordering and absolute mass of neutrinos and GW speed. Finally, the velocity correction to the time delay of signal with arbitrary velocity in general mass profiles is found.

The paper is organized as follows. In Sec. \ref{SecMetric}, we set up the general framework for the calculation of the total travel time $t_\mathrm{if}$ of signal with arbitrary velocity in Schwarzschild spacetime. In Sec. \ref{SecTT}, the $t_\mathrm{if}$  is first evaluated exactly and the result is found as a combination of several elliptic functions. The same  $t_\mathrm{if}$ was then computed in the weak field limit and a much simpler expression is found. Both the exact and approximate $t_\mathrm{if}$ are expressed as functions of signal velocity, source-lens distance, lens-observer distance, angular position of the source and lens mass. In Sec. \ref{SecTD}, time delay $\Delta t_v$ between signals with different velocities but coming from same side of the lens, and time delay $\Delta t_\mathrm{p}$  between signals from different sides of the lens, as well as the difference $\Delta t_{\mathrm{p}v}$ between two $\Delta t_\mathrm{p}$ are found. These time delays are then applied to the cases of supernova neutrino and GW in Sec. \ref{SecApply}. It is shown that the
$\Delta t_v$ might be related to the mass square difference between neutrino mass eigenstates and therefore be used to discriminate neutrino mass orderings, while the difference $\Delta t_{\mathrm{p}v}$ between neutrino and optical signals can be correlated with the absolute mass of neutrinos. For GW, the formula for time delay in a general gravitational potential is derived first. Then the difference $\Delta t_{\mathrm{p}v}$ between GW and GRB is evaluated. It is shown that even for distance as large as $2\times 10^4$ [Mpc] and $v_\mathrm{GW}=(1-10^{-15})c$, $\Delta t_{\mathrm{p}v}$ can only reach to the value of about $1.45\times 10^{-5}$ [s]. To further constrain $v_\mathrm{GW}$ therefore calls for the improvement in uncertainty of GW and GRB time measurements.

\section{Geodesic equations and the total travel time\label{SecMetric}}

We consider the time delay of signal particle/wave to be of different kinds and have different velocities. When passing by a gravitational center, they will experience different travel time. To calculate this travel time, we start from the general spherically symmetric spacetime with metric
\begin{equation}
\dd s^2=A(r)\dd t^2-B(r)\dd r^2-C(r)\lb \dd \theta^2+ \sin^2\theta \dd \varphi^2 \rb \label{SSM}
\end{equation}
where $(t,~r,~\theta,~\phi)$ are the coordinates.
Using the geodesic and normalization equations, we can obtain the following equation of motion in the equatorial plane for coordinate $t$ and $r$
\be
\frac{\dd t}{\dd r}=\frac{E\sqrt{B(r)C(r)}}{\sqrt{A(r)}\sqrt{E^2C(r)-L^2A(r)-\kappa A(r)C(r)}}, \label{dtodr}
\ee
where $\kappa=0,1$ for null and timelike particles respectively. Here $E$ and $L$ are the first integrals of the geodesic equations of $t$ and $\phi$, satisfying
\be
A(r)\frac{\dd t}{\dd \lambda}=E,~C(r)\frac{\dd \varphi}{\dd \lambda}=L.  \label{gdefl}
\ee
They can be interpreted as the energy and orbital angular momentum of the particle per unit mass at infinity. For timelike particles, we have
\be
E=\frac{1}{\sqrt{1-v^2}}, ~L=|\mathbf{p}\times \mathbf{r}|=\frac{v}{\sqrt{1-v^2}}b.\label{lbvrel}
\ee
Here $v$ is the speed of the particle at infinity and $b$ is the impact parameter. For null particles, $E$ would approach infinity but the relation
\be
b=\frac{L}{Ev} \label{bmassive} \ee
holds for both null and timelike particles.

In order to calculate the travel time, in principle we should integrate Eq. \eqref{dtodr} from the source coordinate $r_{\mathrm i}$ to the observer coordinate $r_{\mathrm f}$.
This is usually done by integrating from $r_{\mathrm i}$ to the closest radius $r_0$ first and then adding the same integral from $r_0$ to $r_{\mathrm f}$.
The closest radius $r_0$ is defined as the maximal radial coordinate $r$ satisfying $\displaystyle \frac{\dd r}{\dd t}=0$. Using Eq. \eqref{dtodr} and then substituting $L$ in Eq. \eqref{bmassive}, this is equivalent to
\be
(bEv)^2=\frac{ \lbrack E^2-\kappa A(r_0)\rbrack C(r_0)}{ A(r_0)}, \label{jx0}
\ee
from which $r_0$ can be formally solved in terms of $E$ and $b$.

To facilitate the relevant computation in Schwarzschild spacetime, we now substitute
\be
A(r)=1-\frac{2M}{r},\ B(r)=\left(1-\frac{2M}{r}\right)^{-1},\ C(r)=r^2 \label{SchM}
\ee
into Eq. \eqref{dtodr}  and \eqref{jx0}, and find for timelike particles
\bea
&&\frac{\dd t}{\dd r}=\frac{Er^2}{(r-2M)\sqrt{E^2r^2-L^2(1-\frac{2M}{r})- r (r-2M)}}, \label{dtodrSch1}\\
&&b^2(E^2-1)=\frac{r_0^2 \lsb (E^2-1)r_0+2M\rsb }{r_0-2M }. \label{jx0Sch}
\eea
Note that we use the natural unit $G=c=1$ throughout the paper. The corresponding equations of null particles can be obtained by taking the infinite $E$ limit in the above equations.

For the purpose of later integration, it is convenient to make the change of variable in Eq. \eqref{dtodrSch1} from $r$ to $\displaystyle u=\frac{r_0}{r}$, so that this equation becomes
\begin{equation}
\displaystyle \frac{\dd t}{\dd u}=\frac{Er_0^3}{u^2(2Mu-r_0)\sqrt{r_0\left(2Mu+\lb E^2-1\rb r_0-\frac{u^2(2Mu-r_0)\lbrack 2M+(E^2-1)r_0 \rbrack}{2M-r_0}\right)}}\equiv g(u,r_0,E), \label{dtodrSch}
\end{equation}
where $L$ was replaced using Eq. \eqref{lbvrel} and for simplicity we used $g(u,r_0,E)$ to denote the right hand side.
Then the total travel time $t_\mathrm{if}$  from the source at $r_{\mathrm{i}}$ to the observer at $r_{\mathrm{f}}$ is given by the following integral
\be
t_{\mathrm{if}}
= \int^{\frac{r_0}{r_{\mathrm{i}}}}_1 g(u,r_0,E)\dd u +\int^{\frac{r_0}{r_{\mathrm{f}}}}_1  g(u,r_0,E)\dd u \equiv t_{\mathrm i}+t_{\mathrm f}. \label{eqtif}
\ee
Here the first and second integrals $t_{\mathrm i}$ and $t_{\mathrm f}$ are the times from $r_\mathrm{i}$ to $r_0$ and time from $r_0$ to $r_\mathrm{f}$ respectively. In next section, we will calculate this travel time using  exact integration and the approximation method.

\section{Exact travel time and its approximation \label{SecTT}}

\subsection{Analytical integration}

In order to integrate the Eq. \eqref{eqtif}, we first simplify the integrand $g(u,~r_0,~E)$ to the form
\be
g(u,~r_0,~E)=\frac{c_g}{u^2(u-u_3)\sqrt{(u-1)(u-u_1)(u-u_2)}}, \label{dtodrD}
\ee
where $u_1,\ u_2,\ u_3$ are the roots of the denominator and $c_g$ is a coefficient, given by
\begin{align}
u_{\substack{1\\2}}=&\frac{r_0-2 M}{4 M}\pm \frac{\sqrt{r_0-2M}\sqrt{\left(E^2-1\right) r_0^2+\left(6 E^2-4\right)M r_0-4M^2} }{4M\sqrt{(E^2-1)r_0+2M}},\label{eq:u12def}\\
u_3=&\frac{r_0}{2M},\\
c_g=&\frac{E r_0^3\sqrt{r_0-2M}}{2M\sqrt{2Mr_0\lbr 2M+r_0\lb E^2-1\rb\rbr}}. \label{parameterd}
\end{align}

The time $t_{\mathrm i}$ and $t_{\mathrm f}$ in Eq. \eqref{eqtif} then can be integrated analytically and the result is a linear combination of seven elliptic functions
\be
t_{\mathrm{i/f}}=c_g\sum_{j=0}^7C_j\lsb f_j\left(\frac{r_0}{r_{\mathrm{i/f}}}\right)-f_j(1)\rsb,\label{til}
\ee
where $c_g$ is defined in Eq. \eqref{parameterd} and the coefficients $C_j$'s are
\begin{align}
C_0&=-\frac{1}{u_1 u_2 u_3 },\label{eq:c0def}\\
C_1&=-\frac{\sqrt{u_1-1}}{u_1 u_2 u_3},\\
C_2&=-\frac{u_1 u_3+u_1+u_3-1}{\sqrt{u_1-1} u_1 (u_3-1) u_3},\\
C_3&=\frac{2 \mathrm{i}\  (u_1-1) \left[ u_1^2 (u_2 u_3+u_2+u_3-1)-(u_3-1) (u_1 u_2 u_3+u_1 u_2+u_1 u_3 +u_2 u_3)\right]}{u_1^2 u_2 (u_3-1) u_3 \left(\sqrt{u_1-1}-\sqrt{u_2-1}\right) (u_1-u_3)},\\
C_{\substack{4\\5}}&=\mp \frac{2 \sqrt{u_1-1} \lbrace u_1 \lbrack u_2 (u_3+2)+u_3 \rbrack +u_2 u_3\rbrace }{u_1^2 u_2 u_3^2 \left(\sqrt{u_1-1}-\sqrt{u_2-1}\right)},\\
C_{\substack{6\\7}}&=\mp\frac{4 \mathrm{i}\  \sqrt{u_1-1}}{\sqrt{u_3-1} u_3^2 \left(\sqrt{u_1-1}-\sqrt{u_2-1}\right) (u_1-u_3)},
\end{align}
and the functions $f_j(u)$ are
\begin{align}
f_0(x)=&\frac{\sqrt{(u_1-x) (u_2-x)}}{ x \sqrt{x-1}},\\
f_1(x)=&E\left(\mathrm{i}\ \mathrm{arcsinh}\sqrt{\frac{1-u_1}{x-1}}\bigg|\frac{u_2-1}{u_1-1}\right),\label{f1}\\
f_2(x)=&F\left(\mathrm{i}\ \mathrm{arcsinh}\sqrt{\frac{1-u_1}{x-1}}\bigg|\frac{u_2-1}{u_1-1}\right),\label{f2}\\
f_3(x)=&F\left(\arcsin\sqrt{\frac{h(x)}{h(u_2)}}\bigg|h(u_2)^2\right),\\
f_{\substack{4\\5}}(x)=&\Pi\left(\frac{\mathrm{i}\sqrt{u_1-1}\mp 1}{\mathrm{i}\sqrt{u_1-1}\pm 1}h(u_2);\arcsin\sqrt{\frac{h(x)}{h(u_2)}}\bigg|h(u_2)^2\right),\\
f_{\substack{6\\7}}(x)=&\Pi\left( h(u_2)h(u_3)^{\mp1};\arcsin\sqrt{\frac{h(x)}{h(u_2)}}\bigg|h(u_2)^2\right). \label{eq:f7def} \end{align}
Here ``$\mathrm{i}$'' is the imaginary unit, $F,~E,~\Pi$ are respectively the elliptical integral of the first, second kind and the incomplete elliptic integral defined in Appendix \ref{appda}, and $h(x)$ is an axillary function defined as
\begin{align}
h(x)=\frac{\sqrt{u_1-1}+\sqrt{x-1}}{\sqrt{u_1-1}-\sqrt{x-1}}.
\end{align}

The second term inside the bracket in Eq. \eqref{til} contains the functions $f_j(u)$ evaluated at $u=1$. Evaluation of $f_0(1)$ and $f_1(1)$ however demands special care because they are separately divergent but their divergences  cancel each other exactly and therefore their combination is still finite. For these two terms, we find
\bea
&&\lim_{u\to 1}C_0f_0(u)+C_1f_1(u)\nonumber\\
&=&
C_1\lsb E\lb \frac{\pi}{2}\Big|\frac{u_2-1}{u_1-1}\rb
+\frac{\mathrm{i} \lb u_1-u_2\rb }{\sqrt{\lb u_1-1\rb\lb 1-u_2 \rb}}
F\lb \frac{\pi}{2}\Big|\frac{u_1-1}{u_2-1}\rb
-\mathrm{i} \sqrt{\frac{1-u_2}{u_1-1}}
E\lb \frac{\pi}{2}\Big|\frac{u_1-1}{u_2-1}\rb \rsb
. \label{eqfj1}
\eea
Substituting Eq. \eqref{eqfj1} into \eqref{til} and then \eqref{eqtif}, then the total travel time is found as
\bea
t_{\mathrm{if}}&=&
c_g\left\{\sum_{j=0}^7C_j\lsb f_j\left(\frac{r_0}{r_{\mathrm{i}}}\right)+ f_j\left(\frac{r_0}{r_{\mathrm{f}}}\right) \rsb -2\sum_{j=2}^7C_jf_j(1)-2C_1\lsb E\lb \frac{\pi}{2}\Big|\frac{u_2-1}{u_1-1}\rb\right.\right.\nonumber\\
&& \left.\left.
+\frac{\mathrm{i} \lb u_1-u_2\rb }{\sqrt{\lb u_1-1\rb\lb 1-u_2 \rb}}
F\lb \frac{\pi}{2}\Big|\frac{u_1-1}{u_2-1}\rb
-\mathrm{i} \sqrt{\frac{1-u_2}{u_1-1}}
E\lb \frac{\pi}{2}\Big|\frac{u_1-1}{u_2-1}\rb \rsb\right\}  . \label{tDS}
\eea

\begin{figure}[htp]
\includegraphics[width=0.5\textwidth]{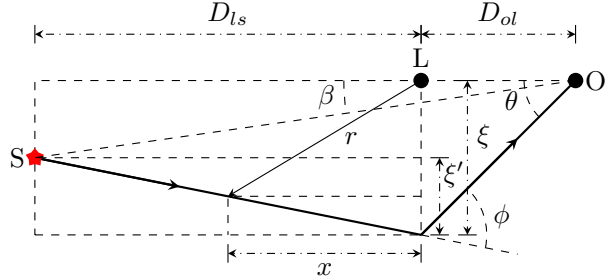}
\caption{Trajectory of particles in GL. S, L and O are the source, lens and observer respectively. $D_{ol}$ and $D_{ls}$ are the distances from observer to lens and from lens to source. $\beta$ is angular position of the source. $\theta$ is the apparent angle of the lensed image. $\phi$ is the deflection angle of the signal. Note that for clarity the other trajectory on the opposite side of the observer-lens axis is not drawn.\label{flensp}}
\end{figure}

Eq. \eqref{tDS} is a function of five parameters $E$ (or $v$), $r_{\mathrm i}$, $r_{\mathrm f}$, $r_0$ and $M$, which once are known, the total travel time would immediately follow. From the definitions of $c_i,~C_i$ and $F_i$ in Eqs. \eqref{eq:u12def}-\eqref{parameterd} and \eqref{eq:c0def}-\eqref{eq:f7def} and dimension counting, one can recognize that if all distance variables $r_{\mathrm i}$, $r_{\mathrm f}$ and $r_0$ are measured in the unit of $M$, then $t_{\mathrm{if}}$ would be linear to $M$, i.e., $t_\mathrm{if}=M\cdot q(E,r_\mathrm{i}/M,r_\mathrm{f}/M)$ for some function $q$. Therefore effectively, the dependence of $t_{\mathrm{if}}$ and the time delays that will be discussed later on $M$ is simple and we can concentrate on dependance on other parameters. Among all parameters $(E, r_\mathrm{i},r_\mathrm{f},r_0,M)$, $r_{\mathrm i},~r_{\mathrm f}$ and $M$ are usually deducible using other astrophysical observations or theoretical tools. $E$ (or $v$) can be measured at the observatory. Therefore, there is only one last obstacle: the closest radius $r_0$, that is not practically known or easily measurable. Although in Eq. \eqref{jx0Sch} we can solve $r_0$ in terms of $E$ and $b$, the impact parameter $b$ is not known explicitly either. One therefore has to find a way to further link $b$ to some measurable quantities in the GL setup. In the remaining part of this section, we will first solve Eq. \eqref{jx0Sch} for $r_0$ and then show that  $b$ can be tied to the apparent angle $\theta$ in the lens equation which is further solvable in terms of $r_\mathrm{i},~r_\mathrm{f}$ and the angular position $\beta$ of the source (see Fig. \ref{flensp}).

From Eq. \eqref{jx0Sch}, which is a cubic polynomial, $r_0$ can be solved as the only positive solution that is accessible by particles coming from and going back to infinity,
\begin{equation}
r_0=\frac{1}{3}\left(-\frac{2M}{E^2-1}+ \frac{\frac{4M^2}{(E^2-1)^2}+3b^2}{f(E,b)}+f(E,b)\right), \label{x0ni}
\end{equation}
where
\begin{align}
f(E,b)=&\bigg\lbrace -\frac{8M^3}{(E^2-1)^3}-\frac{9b^2M(3E^2-2)}{E^2-1}\nonumber\\
&+3\sqrt{3}\sqrt{\frac{16b^2M^4}{(E^2-1)^3}+ \frac{(8-36E^2+27E^4)b^4M^2}{(E^2-1)^2}-b^6}\bigg\rbrace^{1/3}.\label{febdef}
\end{align}
For null rays, this $r_0$ can be simplified to
\be
r_0= \frac{b^2}{\sqrt{3}\lsb b^2\lb -3\sqrt{3}M + \sqrt{27M^2-b^2}\rb \rsb^{1/3}}+\frac{\lsb b^2\lb -3\sqrt{3}M + \sqrt{27M^2-b^2}\rb \rsb^{1/3}}{\sqrt{3}}. \label{dirr0e}
\ee
Eq. \eqref{x0ni} is equivalent to Eq. (14) in Ref. \cite{Jia:2015zon} while Eq. \eqref{dirr0e} agrees with Eq. (6.3.37) in Ref. \cite{Wald:1984}.

Now to connect $b$ to other measurable quantities, we have to use the lens equation. We will only consider this equation in the weak field limit, i.e.,
\be
r_{\mathrm i},~r_{\mathrm f}\gg r_0\gg M\label{eq:x},
\ee although the lens equations in the strong field limit and their solutions are also known \cite{Jia:2015zon}. The reason is that GL in the strong field limit for both the neutrinos and GW are beyond observational capability in the near future. Technically, the lens equation in the strong field limit are also much more involved to solve. The lens equation in the weak field limit is given by (see Fig. \ref{flensp})
\be
\beta = \theta - \frac{r_{\mathrm i}}{r_{\mathrm i}+r_{\mathrm f}}\phi, \label{lenseq}
\ee
where $\beta$ is the angular position of the source, $\theta$ is the apparent angle and $\phi$ is the deflect angle of the geodesic trajectory.
Both $\theta$ and $\phi$ can be linked to the impact parameter $b$. For $\theta$,  its relation with $b$ under the weak field limit is given by
\be
b=r_{\mathrm f}\sin\theta\simeq r_{\mathrm f}\theta. \label{bintheta}
\ee
And for $\phi$ in the weak field limit, its value to the order $\calco(b)^{-1}$ for particles with arbitrary $E$ or $v$ is
\cite{Accioly:2002, Pang:2018}
\be
\phi=\frac{2M\lb 1+\frac{E^2}{E^2-1}\rb }{b}. \label{defang}
\ee
Substituting Eqs. \eqref{bintheta} and \eqref{defang} into \eqref{lenseq}, one obtains a quadratic equation of $b$, whose solutions and the corresponding apparent angles of the images are
\be
b_{\pm}=\frac{r_{\mathrm f}\beta}{2} \pm \frac{1}{2}\sqrt{r_{\mathrm f}^2\beta^2  +\frac{8Mr_{\mathrm i}r_{\mathrm f}}{r_{\mathrm i}+r_{\mathrm f}}\lb 1+\frac{E^2}{E^2-1}\rb },~\theta_\pm=\frac{b_\pm}{r_\mathrm{f}} . \label{impblens}
\ee
The positive $b_+$ corresponds to the particle trajectory along the path on the same side of the lens-observer axis as the source, while the negative $b_-$ corresponds to the path on the other side. Their size satisfies $|b_+|\geq |b_-|$.

Substituting Eq. \eqref{impblens} back into Eq. \eqref{febdef} and further into Eq. \eqref{x0ni}, $r_0$ can be expressed as a function of parameters $E,~\beta,~r_{\mathrm i},~r_{\mathrm f}$ and $M$
\be
r_{0\pm}\equiv r_0\lb E,|b_\pm(E,\beta,r_{\mathrm i},r_{\mathrm f},M)|\rb.\label{r0inb}
\ee
The explicit formula of Eq. \eqref{r0inb} is elementary but too long to show here. It enables the computation of $t_{\mathrm{if}}$ in Eq. \eqref{tDS} in terms of measurables $E~(\mbox{or}~v),~\beta,~r_\mathrm{i},~r_\mathrm{f}$ and $M$.

Because there are two solutions of the impact parameter $b_\pm$ for one set of parameters $( E,~\beta,~r_{\mathrm i},~r_{\mathrm f},~M)$, there are two trajectories connecting the source and the observer and  correspondingly two $t_{\mathrm{if}}$'s. In turn, there  will be two basic types of time delay in the lensing of particles with different velocities: (1) the time delay between the total travel times of particles with different velocities along path on the same side of the lens, and (2) the time delay between total travel times of particles along paths on different sides of the lens. We will denote these two types time delay as $\Delta t_v$ and $\Delta t_\mathrm{p}$ respectively. Of course, if both the particle velocities and path sides are different, the time delay will be a mixed of these two types.

We emphasis that Eqs. \eqref{tDS} and \eqref{x0ni} are {\it exact} formulas for all kinds of lensing including weak, strong or retro- lensings. More importantly, these results are valid for all particle velocity and therefore allow us to study the time delay in lensing of neutrinos and (potentially) massive gravitons. Eqs. \eqref{tDS} do have a drawback that it is expressed using complex elliptical functions which might hinder a simple and clear understanding of the physics, e.g., the effect of various parameters $E$ (or $v$), $\beta$, $r_{\mathrm i}$ and $r_{\mathrm f}$, on the time delay. Therefore, it is desirable to consider an approximation of these results.

\subsection{Approximation in weak field limit}

In the derivation of Eq. \eqref{impblens}, we have used the weak field limit \eqref{eq:x}. In this subsection, we will extend the application of this limit to the integration of Eq. \eqref{dtodrSch} and to the solution \eqref{x0ni}. The key is to note that in this limit, $r_0$ is much larger than $M$.
Therefore, making an asymptotic expansion of small quantity $M/r_0$ in Eq. \eqref{dtodrSch}, it is transformed into
\be
\frac{\dd t}{\dd u}\approx-\frac{E}{u^2\sqrt{(E^2-1)(1-u^2)}}r_0-\frac{EM\lsb-3+2E^2+3u\lb E^2-1\rb\rsb}{(E^2-1)u(u+1)\sqrt{(E^2-1)(1-u^2)}}+\calco\lb \frac{M}{r_0}\rb^1. \label{dtodro1}
\ee
Integrating this using the limits in Eq. \eqref{eqtif} and dropping the terms of order $\calco(M/r_0)$ and higher, the total travel time in the weak field limit becomes
\bea
t_{\mathrm{if,w}}&=
&-\frac{E}{(E^2-1)^{3/2}}\lcb(2E^2-3)M\lcb\ln \lb \frac{r_0^2}{r_{\mathrm i}r_{\mathrm f}}\rb -\ln \lsb\lb 1+\sqrt{1-\frac{r_0^2}{r_{\mathrm f}^2}}\rb \lb 1+\sqrt{1-\frac{r_0^2}{r_{\mathrm i}^2}}\rb \rsb\rcb \right.\nonumber\\
&&\left. -\sqrt{r_{\mathrm f}^2-r_0^2}\lb E^2-1+\frac{E^2M}{r_{\mathrm f}+r_0}\rb-\sqrt{r_{\mathrm i}^2-r_0^2}\lb E^2-1+\frac{E^2M}{r_{\mathrm i}+r_0}\rb\rcb. \label{apprott}
\eea
Beside $M/r_0$, there exists another small ratio $r_0/r_\mathrm{i/f}$ in the weak field limit. Expanding Eq. \eqref{apprott} to the order $\calco(M/r_0)^0$ and $\calco(r_0/r_{i/f})^2$ respectively, one finds
\begin{align}\label{ttexp2}
t_{\mathrm{if,w}}=
& \frac{E M}{\left( E^2-1\right)^{1/2}}\frac{r_{\mathrm f}+r_{\mathrm i}}{r_0}\lb 1-\frac{1}{2}\frac{r_0^2}{r_{\mathrm i} r_{\mathrm f}}\rb \lb \frac{M}{r_0}\rb^{-1} \nonumber\\
&+\frac{ E M }{\left( E^2-1\right)^{3/2} }
\lcb \lsb\left(2E^2-3\right) \ln \left(\frac{4r_{\mathrm i}r_{\mathrm f}}{r_0^2}\right)  +2 E^2\rsb-E^2\frac{  r_{\mathrm f}+r_{\mathrm i} }{r_0}\frac{r_0^2}{r_{\mathrm i} r_{\mathrm f}} \rcb .
\end{align}

Eq. \eqref{ttexp2} and other previous formulas for the total travel time were for particles with energy $E$ of unit mass at infinite radius. Parameter $E$ however is not very convenient for comparing the total travel time for particles with different rest masses. Therefore we replace $E$ by velocity $v$ at infinity using Eq. \eqref{lbvrel} in various formulas. In particular, Eq. \eqref{apprott} becomes
\begin{align}
t_{\mathrm{if,w}}=&\frac{1}{v}\left( \sqrt{r_{\mathrm{i}}^2-r_0^2}+\sqrt{r_{\mathrm{f}} ^2-r_0^2} \right)\nonumber\\
&+\frac{M }{v^3}\lcb (3 v^2 -1) \ln\lsb\left(\frac{\sqrt{r_{\mathrm{i}}^2- r_0^2}+r_{\mathrm{i}}}{r_0}\right)\left(\frac{\sqrt{r_{\mathrm{f}} ^2- r_0^2}+r_{\mathrm{f}} }{r_0}\right)\rsb+\sqrt{\frac{r_{\mathrm{i}}- r_0}{r_{\mathrm{i}}+ r_0}}+\sqrt{\frac{r_{\mathrm{f}} - r_0}{r_{\mathrm{f}} + r_0}} \rcb. \label{approttv}
\end{align}
The first term is of geometrical origin and represents the propagation time for particle with general velocity $v$ along the bent path. The second term represent the effect of the general relativistic gravitational potential to the total travel time.
When $v=c$, this becomes the well-known total travel time for null particles \cite{Hartle:2003yu}.
For Eq. \eqref{ttexp2}, after substitution of $v$, the total travel time is approximated by
\begin{align}
t_{\mathrm{if,w}}=\frac{r_{\mathrm{i}} +r_{\mathrm{f}}}{v}-\frac{ r_0^2( r_{\mathrm{i}} +r_{\mathrm{f}} )}{2 v r_{\mathrm{i}}  r_{\mathrm{f}} }
+\frac{M}{v^3} \lsb\left(3 v^2 -1\right) \ln  \left(\frac{4r_{\mathrm{i}} r_{\mathrm{f}}}{r_0^2}\right)  +2 \rsb -\frac{M r_0( r_{\mathrm{i}} +r_{\mathrm{f}} )}{r_{\mathrm{i}}  r_{\mathrm{f}} v^3}. \label{ttbigyzv}
\end{align}
The first term is the time cost for travel if the spacetime is Minkovski and the second term is the correction because the bending of the geodesic trajectory causes extra distance, and both these two terms originate from the geometric propagation time term in Eq. \eqref{approttv}. The third term in Eq. \eqref{ttbigyzv}, when setting $v=c$, is half of the conventional Shapiro time delay for a returning light signal. The last term is the high order term from the general relativistic potential term in Eq. \eqref{approttv}. Note that although we have used the weak field limit  \eqref{eq:x} to various orders in the derivation of Eqs. \eqref{approttv} and \eqref{ttbigyzv}, no assumption on velocity $v$ was used and therefore Eqs. \eqref{approttv} and \eqref{ttbigyzv} should be valid for any velocity.

The $r_0$ in Eq. \eqref{ttexp2}, which was given by Eq. \eqref{r0inb},
can also be further simplified in the weak field limit. The key is that this limit not only implies $r_0\gg M$ but also $|b_\pm|\gg M$. Therefore expanding the right side of solution \eqref{r0inb} for large $b$, one obtains to the $\calco(b_\pm)^0$ order the following result
\be
r_{0\pm\mathrm{w}}= |b_\pm|-\frac{E^2M}{E^2-1}+\calco( b_\pm )^{-1},
 \label{impbo1}
\ee
which after substituting Eq. \eqref{impblens} for $b_\pm$ and $v$ for $E$ becomes
\begin{align}
r_{0\pm\mathrm{w}}
\approx &\frac{1}{2}\sqrt{r_{\mathrm{f}}^2 \beta^2 +
\frac{8Mr_{\mathrm i}r_{\mathrm f}}{r_{\mathrm i}+r_{\mathrm f}}\lb 1+\frac{1}{v^2}\rb } \pm\frac{r_{\mathrm{f}}\beta}{2}-\frac{M}{v^2}. \label{x0ap}
\end{align}
Substituting this into Eq. \eqref{approttv} or \eqref{ttbigyzv}, one then can obtain the total time for each set of parameters in a much simpler way than Eqs.
\eqref{tDS} and \eqref{r0inb}. To check the correctness of these results, we have numerically calculated the $t_\mathrm{if}$'s using both Eqs. \eqref{tDS} and \eqref{ttbigyzv} with variables within the parameter ranges used in Sec. \ref{SecTD} and \ref{SecApply} and excellent agreement was found.

\section{Two types of time delay\label{SecTD}}

As we pointed out in the previous section, there will be two basic types of time delay if signals with different velocities are lensed: the time delay $\Delta t_v$ between particles of different velocities, and the time delay $\Delta t_\mathrm{p}$ between particles traveled along different paths.
In principle we can use both the exact result Eq. \eqref{tDS} and the weak field limit result Eq. \eqref{approttv} to find all these time delays. However, for the purpose of later usage and more intuitive understanding of relevant results, in this section we will derive perturbative results for these two types of time delay starting from Eq. \eqref{ttbigyzv}. Note that the total travel time $t_{\mathrm{if,w}}$ is dependent on variables $v,~r_{\mathrm{i}},~r_{\mathrm{f}},~M$ explicitly and on $\beta$ and the path choice ($\pm$ sign) implicitly through $r_{0\mathrm{w}\pm}$ in Eq. \eqref{x0ap}. In other words,
\be
t_{\mathrm{if,w}}=t_{\mathrm{if,w}}\lb v,r_0\lb v,|b_\pm(v,\beta,r_{\mathrm i},r_{\mathrm f},M)|\rb,r_{\mathrm{i}},r_{\mathrm{f}},M\rb.\ee
In the computations below, for the simplicity of the notation, we will only keep necessary variables and suppress the rest.

\subsection{Time delay $\Delta t_v$\label{subsectdv}}

We consider the situation that two signals with velocities $v$ and $v^\prime~(v^\prime>v)$ traveling on same side of the lens first. If $v$ and $v^\prime$ are very different, then the time delay should be evaluated directly using
Eq. \eqref{ttbigyzv}
\be
\Delta t_v=t_{\mathrm{if,w}} (v)-t_{\mathrm{if,w}}(v^\prime). \label{eqdtvdef}
\ee
Given that for all practical $\beta$, the weak field limit \eqref{eq:x} implies that the main contribution to $\Delta t_v$ comes from the first term in Eq. \eqref{ttbigyzv}. In other words, we should roughly have
\be \Delta t_v\approx (r_\mathrm{i}+r_\mathrm{f})\left(\frac{1}{v}-\frac{1}{v^\prime}\right) +~\mathrm{high~order~terms~containing~}r_0. \label{eqdtvapp}\ee
Since the dependance of $\Delta t_v$ on $\beta$ is hidden in $r_0$, this approximation not only fixes the main dependance of $\Delta t_v$ on the coordinates and velocities, but also suggest that the time delay in this case is largely insensitive to the angular position $\beta$ of the source. For the high order terms in Eq. \eqref{eqdtvapp}, from Eq. \eqref{ttbigyzv} one can see that they will be maximal when $r_0$ is large while $r_\mathrm{i}$ and $r_\mathrm{f}$ is relatively small. Expansion \eqref{x0ap} further suggests that $r_0$ is large only when $\beta$ is large. Therefore the dependance of $\Delta t_v$ on $\beta$ is stronger when $\beta$ is large and less so when it is small.

If $v$ and $v^\prime$ are very close so that their difference $\Delta v=v^\prime-v$ is much smaller than $v^\prime$ and $v$, e.g. for a null ray and an ultra-relativistic ray or two ultra-relativistic rays, then a further expansion of Eq. \eqref{eqdtvdef} around $v$ can be carried out to find
\be
\Delta t_v\approx -\frac{\dd t_{\mathrm{if,w}} (v)}{\dd v}\Delta v +\calco(\Delta v)^2, \label{deltat}
\ee
where
\begin{align}
\frac{\dd t_{\mathrm{if,w}}}{\dd v}=&-\frac{r_{\mathrm{i}} +r_{\mathrm{f}}}{v^2}+\frac{\lb r_{\mathrm{i}} +r_{\mathrm{f}}\rb r_{0\pm\mathrm{w}} \lb r_{0\pm\mathrm{w}}-2 v \frac{\dd r_{0\pm\mathrm{w}}}{\dd v} \rb }{2 v^2 r_{\mathrm{i}}  r_{\mathrm{f}} }\nonumber\\
&-\frac{ M\lcb 3 r_{0\pm\mathrm{w}} \lsb\left( v^2 -1\right) \ln  \left(\frac{4r_{\mathrm{f}} r_{\mathrm{i}}}{r_{0\pm\mathrm{w}}^2}\right)  +2 \rsb +2 v \lb 3 v^2 -1\rb\frac{\dd r_{0\pm\mathrm{w}}}{\dd v} \rcb}{r_{0\pm\mathrm{w}} v^4}\nonumber\\
&+\frac{M \lb r_{\mathrm{i}} +r_{\mathrm{f}} \rb \lb 3 r_{0\pm\mathrm{w}}- v \frac{\dd r_{0\pm\mathrm{w}}}{\dd v}\rb}{r_{\mathrm{i}}  r_{\mathrm{f}} v^4}. \label{dtode}
\end{align}
and $r_{0\pm\mathrm{w}}$ was still given by Eq. \eqref{x0ap} and consequently
\be
\frac{\dd r_{0\pm\mathrm{w}}}{\dd v}=\frac{2 M}{v^3}\lb 1-\frac{2 r_{\mathrm{i}} \sqrt{r_{\mathrm{f}} }}{\sqrt{r_{\mathrm{i}}+r_{\mathrm{f}} } \sqrt{\beta ^2 r_{\mathrm{f}} \lb r_{\mathrm{i}}+ r_{\mathrm{f}}  \rb +8 r_{\mathrm{i}} M \lb 1+ \frac{1}{v^2}\rb}} \rb. \label{drodeap}
\ee

\begin{center}
\begin{figure}
\includegraphics[width=0.49\textwidth]{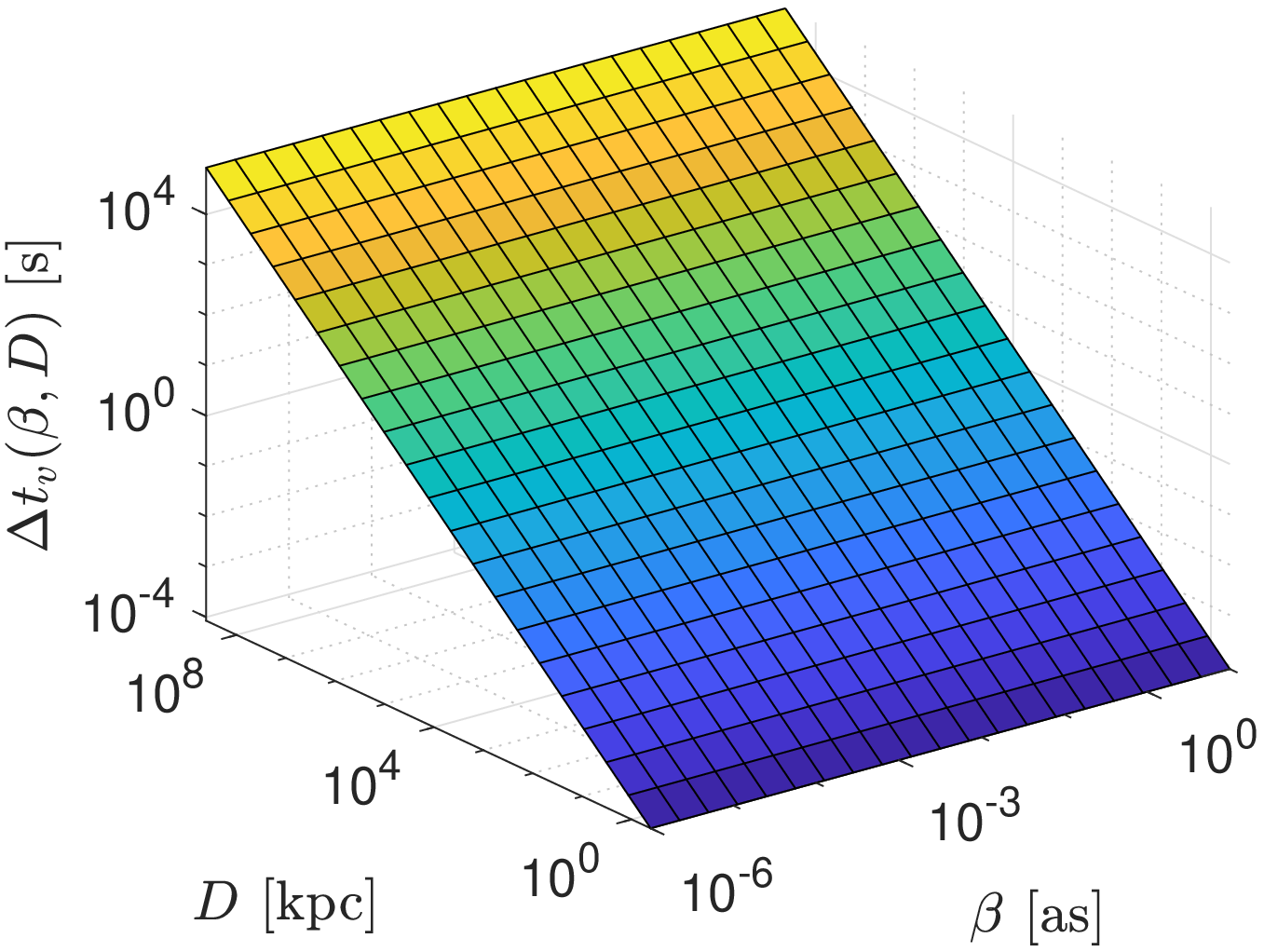}
\includegraphics[width=0.49\textwidth]{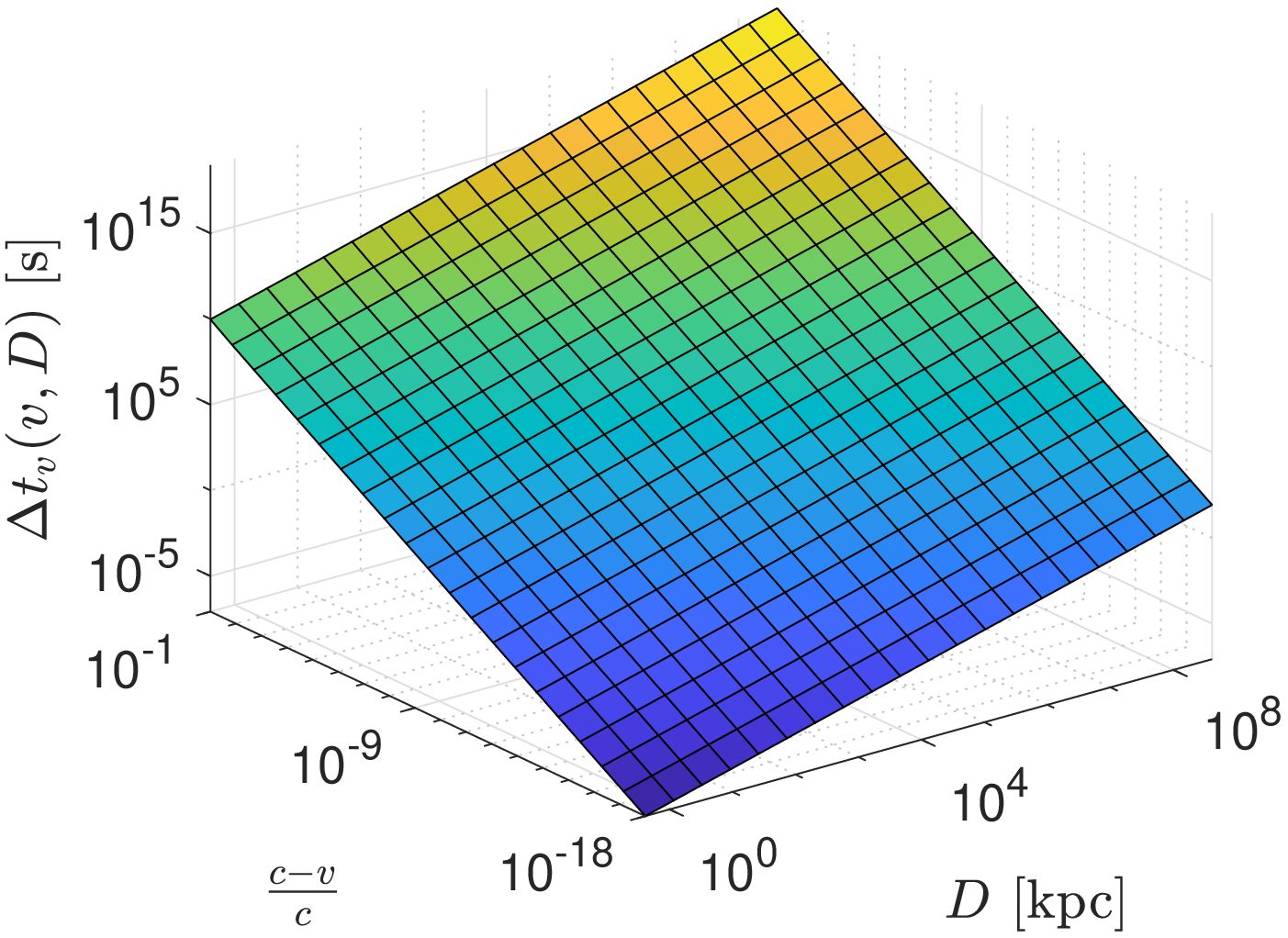}\\
(a)\hspace{6cm}(b)
\caption{\label{figtdofsameside} Time delay $\Delta t_v$ between signals of different velocity on the same sides of the lens axis (a) as a function of $\beta$ and $D$ for $v=(1-10^{-15})c$, and (b) as a function of $D$ and $v$ for $\beta=1$ [as]. }
\end{figure}
\end{center}

To reveal more physical insights from these results, we plot in Fig. \ref{figtdofsameside} the time delay $\Delta t_v$ using the definition \eqref{eqdtvdef} and total time formula Eq. \eqref{ttbigyzv}.
Note the $\Delta t_v$ depends on four variables nontrivially: $\beta, ~v$ and $r_{\mathrm f}/M$ and $r_{\mathrm i}/M$. The typical range of $\beta$ is from $10^{-6}$ [as] to 10 [as] and $v/c$ is between 0 and 1. For $r_{\mathrm i}$ and $r_{\mathrm f}$, for simplicity we fix them to be equal and denote them collectively by $D$. It can range from $(1\sim 10~\mathrm{[kpc]})/(10^{-1}\sim 10~[M_\odot])$ in the typical microlensing case to $(1\sim 10^2~\mathrm{[Mpc]})/(10^6\sim 10^{10} ~ M_\odot)$ if the lensing is due to supermassive black holes (except the Galatic one), to even
$(10^2\sim 10^5~\mathrm{[Mpc]})/(10^{10}\sim 10^{12}~M_\odot)$ for typical lensing by galaxies. Therefore we will use a range of $(10^{-7}\sim 10^2~ \mathrm{[kpc]})/M_\odot$ for $D/M$. The value of $M$ is implicitly fixed at $M=4.12\times 10^6 M_\odot$ in this plot. $\Delta t_v$ at other values of $M$ can be obtained by scalings.

In Fig. \ref{figtdofsameside} (a), the time delay $\Delta  t_v$ between signals with velocity $v=(1-10^{-15})c$ and light as a function of $\beta$ and $D$ is plotted. This velocity $v$ is chosen to represent an allowed value of the GW velocity \cite{GBM:2017lvd,Monitor:2017mdv}. It is seen that as discussed in Eq. \eqref{eqdtvapp}, the dependance of the time delay on $D$ is linear, while its dependance on $\beta$ is not noticeable in this plot. For $D= 8.02$ [kpc], a source located at the same distance from the galaxy center as Earth and then lensed by the galactic supermassive BH, the $\Delta t_v$ for signal with this velocity is about 0.0020 [s]. This is more than one order smaller than the 0.054 [s] uncertainty in the measurement of time delay between GW and GRB \cite{GBM:2017lvd,Monitor:2017mdv} and therefore calling for improvement in GW and GRB detection time accuracy if $\Delta t_v$ in lensing of galactic central BH for galactic merger event is ever used.
In Fig. \ref{figtdofsameside} (b), the time delay $\Delta t_v$ between light and signal with arbitrary velocity $v$ is plotted as a function of $v$ and $D$ for $\beta=1$ [as]. This more clearly verified the observation in Eq. \eqref{eqdtvapp} that $\Delta t_v$ is not only linear to the distance but also linear to
$ \left(\frac{1}{v}-\frac{1}{v^\prime}\right)=\left(\frac{1}{1-\Delta v}-1\right)\approx \Delta v$.
It is seen that when the velocity difference decreases from $v=(1-10^{-14})c$ to $v=(1-10^{-16})c$, the time delay also decreases linearly by two orders for all $D$.
We also numerically verified that, if $\beta$ is changed to larger value (e.g. 10 [as]) or smaller value (e.g. $10^{-6}$ [as]), the change in plot Fig. \ref{figtdofsameside} (b) is indeed unnoticeable, in agreement with previous observation.

\subsection{Time delay $\Delta t_\mathrm{p}$\label{sectdtwop}}

For particles traveling on two sides of the lens along the $b_-$ and $b_+$ paths respectively, the time delay can be expressed using Eq. \eqref{ttbigyzv} as
\be
\Delta t_{\mathrm p}= t_{\mathrm{if,w}}\lb v_-,r_0( v_-,|b_-(v_-)|)\rb -
t_{\mathrm{if,w}}\lb v_+,r_0( v_+,|b_+(v_-)|)\rb, \label{dtb}
\ee
where $v_\pm$ are the velocity of the signal on two sides of the lens and  $r_0(v_\pm,|b_\pm(v_\pm)|)$ are given by Eq. \eqref{r0inb}.
This equation is valid for comparing any two kinds of signal with arbitrary velocities.

In astrophysical observation, it is often the case that total travel time of signals  with same velocities from different lensing images are compared. In this situation, $v_+=v_-=v$ and
the time delay Eq. \eqref{dtb} becomes after using Eq. \eqref{ttbigyzv}
\begin{align}
\Delta t_\mathrm{p}(v)
\approx \frac{\lb r_{\mathrm{f}}+r_{\mathrm{i}}\rb \lb r_{0+\mathrm{w}}^2 - r_{0-\mathrm{w}}^2\rb }{2 r_{\mathrm{f}} r_{\mathrm{i}} v} + \frac{2 M \lb 3 v^2 -1\rb \lb \ln r_{0+\mathrm{w}} - \ln r_{0-\mathrm{w}}\rb}{v^3}, \label{eq:675}
\end{align}
where we have ignored the last term in Eq. \eqref{ttbigyzv}, which is valid when $v$ is not extremely small: $(v^2/c^2>M/r_{0\pm \mathrm{w}})$.

When $\beta$ is small, the $|r_\mathrm{0+w}|-|r_\mathrm{0-w}|\propto \beta^1$ will also be small and therefore the perturbative expansion of Eq. \eqref{eq:675} in powers of $\beta$ and then in powers of $M/r_\mathrm{i/f}$ can be done. To the $\beta^3$ order and leading order of $M/r_\mathrm{i/f}$ in each order of $\beta^i$, one finds
\begin{align}
\Delta t_{\mathrm p}\approx& 4 \sqrt{2} M \lsb\frac{r_{\mathrm{f}}\lb r_{\mathrm{f}} + r_{\mathrm{i}}\rb}{r_{\mathrm{i}} M \lb 1+v^2\rb}\rsb^{1/2} \beta + \frac{M }{6 \sqrt{2}} \lsb \frac{r_{\mathrm{f}}\lb r_{\mathrm{f}} + r_{\mathrm{i}}\rb}{r_{\mathrm{i}} M \lb 1 + v^2\rb}\rsb^{3/2}\beta^3 +\calco\lb \beta\rb^5. \label{tdbap2}
\end{align}
In order for this expansion to converge, then we should demand that
\be
\beta \ll \sqrt{\frac{r_{\mathrm i}M(1+v^2)}
{r_{\mathrm f}\lb r_{\mathrm i}+r_{\mathrm f}\rb}}. \label{betacond}
\ee
If $\beta$ is in this range, then clearly the first term will dominate and therefore it is expected that the time delay $\Delta t_\mathrm{p}$ will be proportional to $\beta^1$ and $r_\mathrm{i/f}^{1/2}$.
Note however for some gravitational lensing  with large $\beta$ and/or $r_\mathrm{i/f}$, this condition is violated (see Ref. \cite{Treu:2010uj} for ranges of parameters $\beta$ and $r_\mathrm{i/f}$) and therefore for those cases the expansion \eqref{tdbap2} will not be accurate. In those cases, one can easily show that the logarithmic term of Eq. \eqref{eq:675} will be much smaller than its first term, which after substituting Eq. \eqref{x0ap} for $r_{0\pm\mathrm{w}}$ becomes
\begin{align}
\Delta t_\mathrm{p}=&-\frac{\beta \sqrt{r_\mathrm{f}(r_\mathrm{i}+r_\mathrm{f}) \left[\beta^2 r_\mathrm{f}(r_\mathrm{i}+r_\mathrm{f})+8M r_\mathrm{i}\left(1+\frac{1}{v^2}\right)\right]}}{2 r_\mathrm{f}v}+\mathrm{~ln~terms}.
\label{dtbf}
\end{align}
Note that when the relativistic limit of the velocity $v$ is taken, the $\calco(1-v)^1$ order term in Eq. \eqref{dtbf} agrees with Eq. (20) of Ref. \cite{Glicenstein:2017lrm}. If we further take the limit of large $r_\mathrm{i}$ and $ r_\mathrm{f}$, this becomes
\begin{align}
\Delta t_\mathrm{p}
\approx&\frac{(r_\mathrm{i}+r_\mathrm{f}) \beta ^2 }{2r_\mathrm{f} v}+\mathrm{~high~order~terms}. \label{eq:ldtpexp}
\end{align}
Unlike the situation in Eq. \eqref{tdbap2},  this time delay is proportional to $\beta^2$ and $r_\mathrm{i}^1$.

Regarding the dependance of $\Delta t_\mathrm{p}$ on the signal velocity $v$ that is close to $c$, in both cases of Eq. \eqref{tdbap2} and \eqref{eq:ldtpexp},  a further expansion of their first term implies that the dominate part of $\Delta t_\mathrm{p}$ is always proportional to $ 1+\Delta v$. This is very close to 1 for relativistic particles, suggesting that in this case the variation of the time delay due to velocity change is much smaller than the time delay itself.

\begin{center}
\begin{figure}[htp!]
\includegraphics[width=0.49\textwidth]{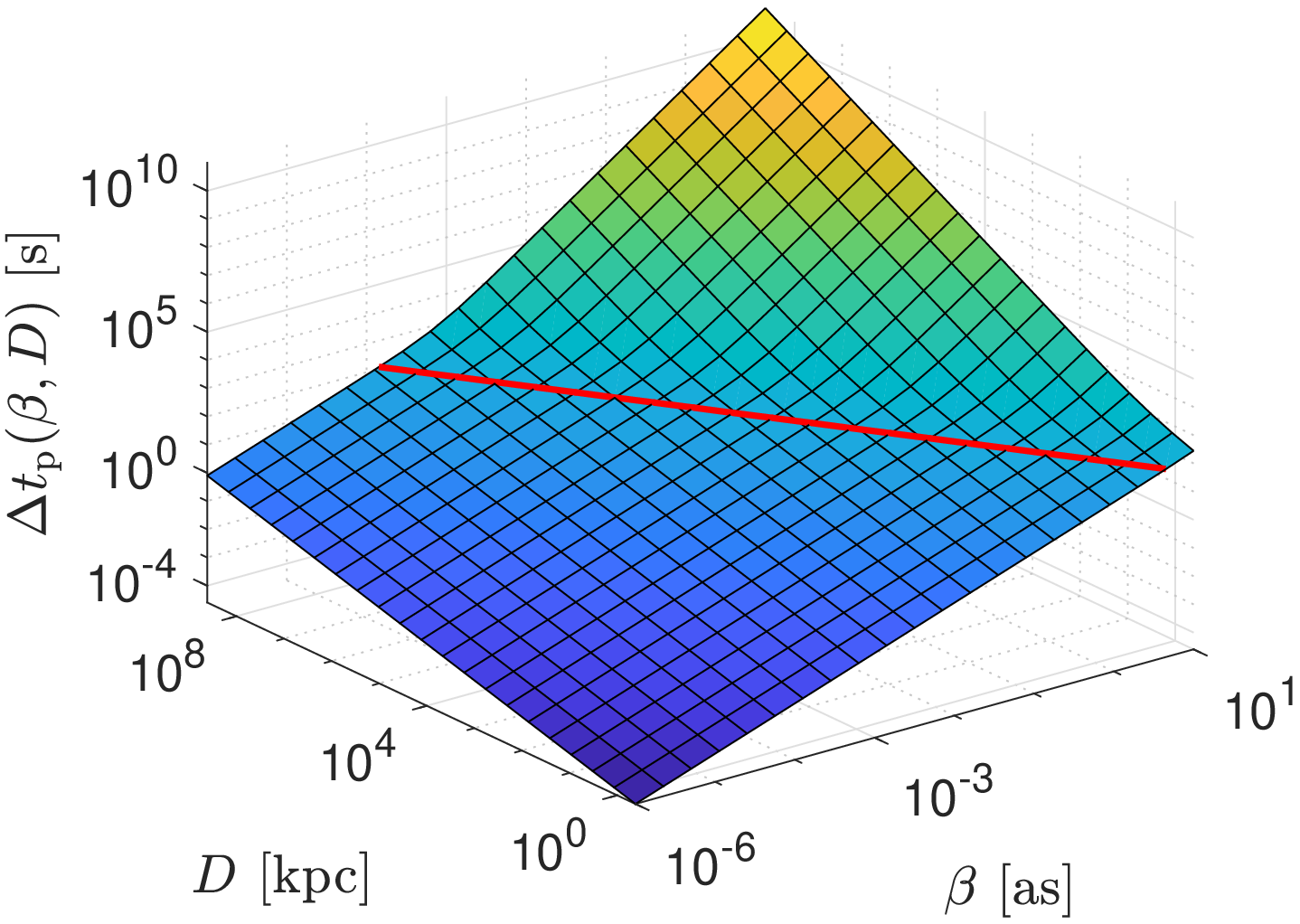}
\includegraphics[width=0.49\textwidth]{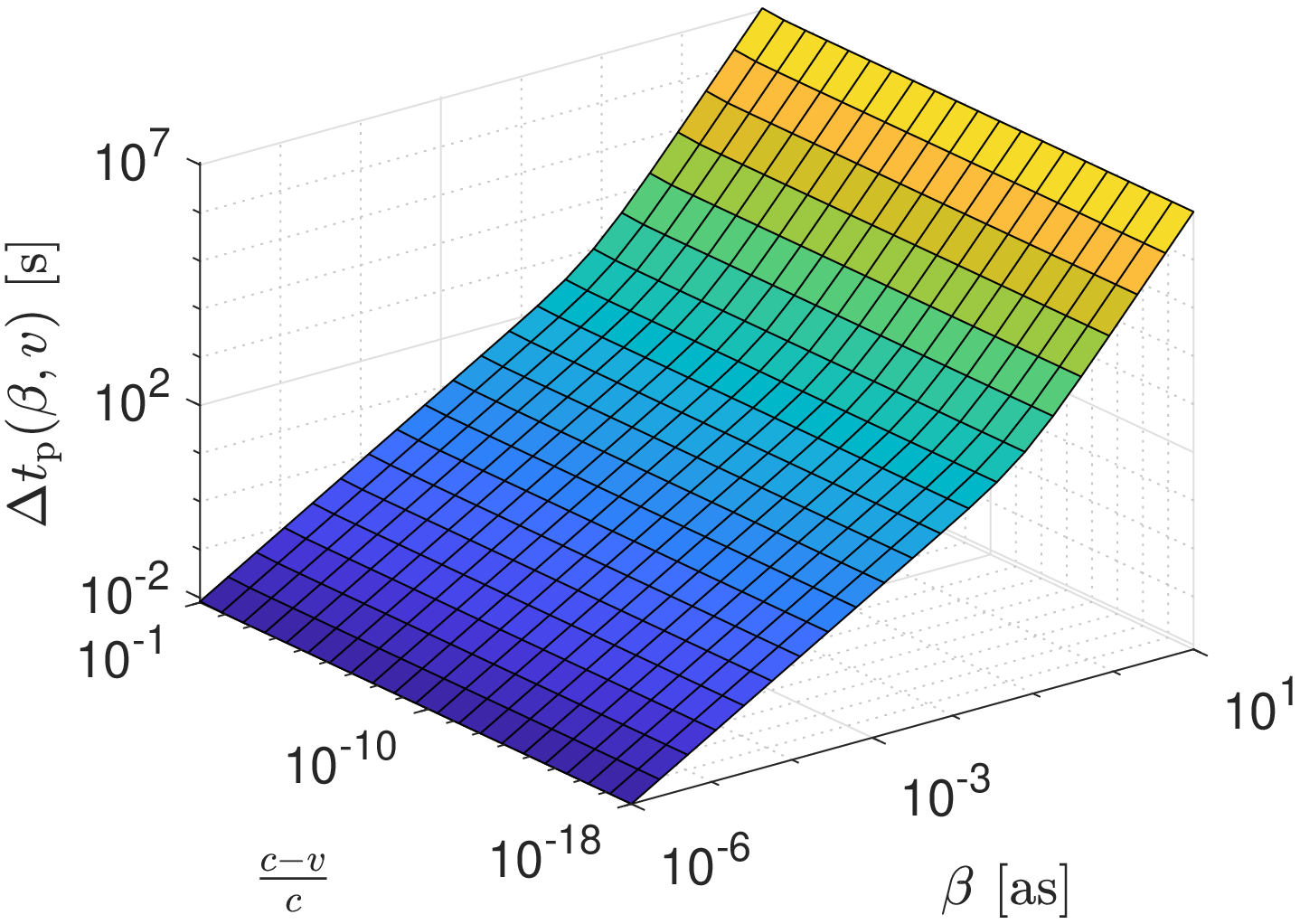}\\
(a)\hspace{6cm}(b)
\caption{\label{figtdofm} Time delay $\Delta t_\mathrm{p}$: (a) as a function of $\beta$ and $D$ for fixed $v=c$; and (b) as a function of $\beta$ and $v$ for fixed $D=4.12\times 10^1$ [Mpc]. The line in (a) is drawn according to Eq. \eqref{betacondapp}.}
\end{figure}
\end{center}

In Fig. \ref{figtdofm}, we plot the time delay $\Delta t_\mathrm{p}$ as functions of various parameters using Eq. \eqref{eq:675}.
In Fig. \ref{figtdofm} (a), $\Delta t_\mathrm{p}$ as a function of $\beta$ and $r_\mathrm{i}=r_\mathrm{f}=D$ is plotted for light signal $(v=c)$ using logarithmic scale. It is seen that for fixed $D$ and increasing $\beta$, $\Delta t_\mathrm{p}$ increases linearly with slope 1 in the log-scale plot when  $\beta$ is small. Similarly, for fixed $\beta$ and increasing $D$, $\Delta t_\mathrm{p}$ also increases linearly but with slope 1/2 when $D$ is small. These features are in agreement with Eq. \eqref{tdbap2}.
When Eq. \eqref{betacond} is about to be violated as $\beta$ and $D$ increase, i.e., when
\be
\beta \approx \sqrt{\frac{r_{\mathrm i}M(1+v^2)}
{r_{\mathrm f}\lb r_{\mathrm i}+r_{\mathrm f}\rb}} \label{betacondapp}
\ee
a transition from Eq. \eqref{tdbap2} to \eqref{eq:ldtpexp} happens. This  can be seen from the coincidence of the red line representing Eq. \eqref{betacondapp} and the bending region in Fig. \ref{figtdofm} (a). Beyond this line, the slope
of the plot in both the $\beta$ and $D$ directions are doubled, reflecting that Eq. \eqref{eq:ldtpexp} now takes place.

In Fig. \ref{figtdofm} (b), $\Delta t_\mathrm{p}$ as a function of $\beta$ and $v$ is present.
It is seen that for the range of $\beta$, the same transition from small $\beta$ expansion Eq. \eqref{tdbap2} to large $\beta$ expansion Eq. \eqref{eq:ldtpexp} happens. Moreover, for the entire parameter range, the dependance of $\Delta t_\mathrm{p}$ on $v$ is not noticeable in this log-scale plot, as argued below Eq. \eqref{eq:ldtpexp}. Indeed, one can plot $\Delta t_\mathrm{p}$ as function of $v$ and $D$ too and also find the very weak dependance on $v$.

In Sec. \ref{SecApply}, we will be interested in the difference of two $\Delta t_\mathrm{p}$'s of different velocities $v$ and $v^\prime$, i.e., $ \Delta t_{\mathrm{p}v}\equiv \Delta t_\mathrm{p}(v ) -\Delta t_\mathrm{p}(v^\prime)$, which can be calculated using Eq. \eqref{eq:675}. In the entire parameter ranges of $M$, $\beta$ and $r_\mathrm{i/f}$ considered in Fig. \ref{figtdofm} (a), one can verify that when both $v^\prime$ and $v$ are close to $c$, the contribution from the $M/v^2$ term in $r_{0\pm\mathrm{w}}$ in Eq. \eqref{x0ap} to this difference can be ignored. Further expand this difference to the first order of $(v^\prime -v)$, the result is found as
\begin{align}
 \Delta t_{\mathrm{p}v}
\approx &\frac{\beta \sqrt{ r_\mathrm{f}(r_\mathrm{i}+r_\mathrm{f}) \left[\beta^2 r_\mathrm{f}(r_\mathrm{i}+r_\mathrm{f})+16M r_\mathrm{i}\right]}}{2r_\mathrm{i}}( v^\prime -v). \label{eqdtpv}
\end{align}
Similar to the situation in Eqs. \eqref{tdbap2} and \eqref{eq:ldtpexp}, the coefficient of $( v^\prime -v)$ in Eq. \eqref{eqdtpv} also depends on $\beta$ and $r_\mathrm{i/f}$ as $\beta^1 r_\mathrm{i/f}^{1/2}$ when they are small and as $\beta^2 r_\mathrm{i/f}^1$ when they are large. Later in Sec. \ref{SecApply} we will use this to find the time delay difference between two neutrino mass eigenstates in the neutrino lensing case and between GRB and GW signals in the binary neutron star merger case.

\section{Time delay of neutrinos and GWs\label{SecApply}}

\subsection{SNN time delay}

The neutrino mass ordering and the absolute value of neutrino masses are important problems for not only particle physics but also cosmology and astrophysics. The latest constraints on the neutrino mass square differences are \cite{Tanabashi:2018oca}
\bea
&&\Delta m_{21}^2=(7.53\pm 0.18)\times 10^{-5} \mbox{eV}^2,\label{nmdiff1}\\
&&\Delta m_{32}^2=(-2.56\pm 0.04)\times 10^{-3} \mbox{eV}^2~(\mbox{inverted ordering}),\\
&&\Delta m_{32}^2=(2.51\pm 0.05)\times 10^{-3} \mbox{eV}^2~(\mbox{normal ordering}),
\label{nmdiff}
\eea
and the cosmological bound on the sum of the neutrino masses is \cite{Ade:2015xua}
\be \sum_j m_j<0.170~\mbox{eV}. \label{nmsum} \ee

Using the Eqs. \eqref{nmdiff1} to \eqref{nmsum}, we can estimate the masses of neutrinos for both the normal and inverted orderings.
Assuming the lightest neutrino is massless, Eq. \eqref{nmdiff1} to \eqref{nmdiff} suggests that for normal order
\be m_1\approx0~\mbox{eV}, ~m_2\approx 8.678\times 10^{-3}~\mbox{eV}, ~m_3\approx 5.085\times 10^{-2}~\mbox{eV} \label{nmno1}\ee
and for inverted order
\be m_1\approx 4.985\times 10^{-2}~\mbox{eV},~ m_2\approx 5.060\times 10^{-2}~\mbox{eV}, ~m_3\approx 0~\mbox{eV}. \label{nmio1}\ee
If we assume that the bound \eqref{nmsum} is saturated, then the corresponding neutrino masses for normal order are
\be m_1\approx4.923\times 10^{-2}~\mbox{eV} , ~m_2\approx4.999\times 10^{-2}~\mbox{eV} , ~m_3\approx 7.078\times 10^{-2}~\mbox{eV} \label{nmno2}\ee
and for inverted order
\be m_1\approx 6.436\times 10^{-2}~\mbox{eV},~ m_2\approx 6.494\times 10^{-2}~\mbox{eV},~ m_3\approx 4.071\times 10^{-2}~\mbox{eV}. \label{nmio2}\ee

We first consider the time delay of SNN signals from the same side of the lens. In this work, we focus on the SNNs because their properties are better understood comparing to neutrinos of other astrophysical origin \cite{Aartsen:2016ngq,IceCube:2018cha,IceCube:2018dnn}.
Because neutrinos have three mass eigenstates, for any given SNN spectrum that usually last a few seconds the three mass eigenstates will decouple from each other during propagation from supernova to observer which costs long time. Therefore it is expected that three separate signals corresponding to the $|\nu_1\rangle$, $|\nu_2\rangle$ and $|\nu_3\rangle$ eigenstates will be received from the same side of the lens. Here we will show that these three signals will have a time delay that might be used to resolve the neutrino mass ordering problem.

\begin{center}
\begin{figure}[htp!]
\includegraphics[width=0.6\textwidth]{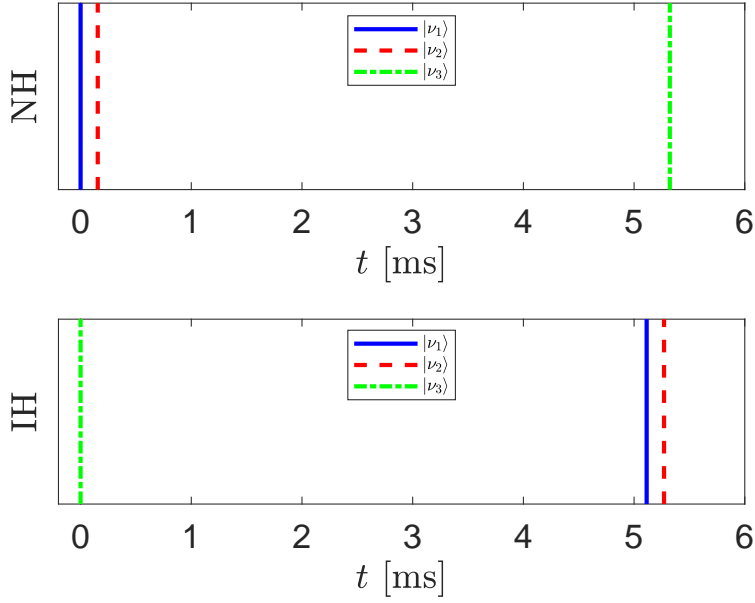}
\caption{\label{figoneside1} Time delay $\Delta t_{v,ij}$ between different neutrino mass eigenstates $|\nu_i\rangle$ from the same side images. The parameters in the lensing model are $M=10^{11} M_{\odot}$, $r_\mathrm{i}=r_\mathrm{f}=2$ [Mpc], $\beta=1$ [as] and neutrino energy $E=10$ MeV. The firstly arrived signal time were set to $t=0$.}
\end{figure}
\end{center}

We assume that the SNN has a fixed energy of 10 MeV which is about the average of the spectrum \cite{Hirata:1987hu, Bionta:1987qt}. Using the masses in Eqs. \eqref{nmno1} to \eqref{nmio2}, we then can calculate the velocity $v_i,~v_j$ of each mass eigenstate $|\nu_i\rangle$ and $|\nu_j\rangle$ respectively and use Eq. \eqref{eqdtvdef} to find the time delay $\Delta t_{v,ij}$ between them, i.e., $\Delta t_{v,ij}\equiv t_\mathrm{if,w}(v_i)-t_\mathrm{if,w}(v_j)$, for both mass orderings. For simplicity, we assume that the lens is located at 2 [Mpc] away from both the observer and the source and the source angular position $\beta=1$ [as]. In Fig. \ref{figoneside1}, we show the time delays of both the normal and inverted orderings with the masses given by Eqs. \eqref{nmno1}-\eqref{nmio1}.
It is seen that for the normal ordering, there is a delay of about 0.1 [ms] for $|\nu_2\rangle$ signal comparing to $|\nu_1\rangle$ and about 2.6 [ms] for $|\nu_3\rangle$ comparing to $|\nu_2\rangle$. For the inverted ordering,  $|\nu_1\rangle$ appears 2.6 [ms] after the $|\nu_3\rangle$ and $|\nu_2\rangle$ appears 0.1 [ms] after $|\nu_1\rangle$. In both the mass orderings, if $|\nu_1\rangle$ and $|\nu_2\rangle$ signal are to be resolved, then these signals should have a characteristic time that is narrower than 0.1 [ms]. Fortunately for supernovae that collapse into BHs, it is known that the SNN spectrum tail has a characteristic termination time that last usually about $2R/c\approx 0.1$ [ms], where $R\approx 10$ [km] is the size of neutrino emission region in supernova. If the location of the lens and source are 30 time larger, then it can be seen through Eq. \eqref{eqdtvapp} that the time intervals between the mass eigenstates will be 30 larger too. That is, $|\nu_1\rangle$ and $|\nu_2\rangle$ are separated by about 3 [ms]. This is about the minimal time duration of another feature that is widely believed to exist in SNN spectrum, the neutronization burst peak \cite{Hempel:2011mk,Lentz:2011aa,Bruenn:2012mj,Couch:2013kma}. Therefore in this case the mass eigenstates from the neutronization burst peaks might also be split in time in different ways in these two mass orderings. These all suggest that the two mass orderings will appear as two different sequence of events separated by different time intervals if the distance of the lens and supernova are large enough.

We also varied the mass from Eqs. \eqref{nmno1}-\eqref{nmio1} to \eqref{nmno2}-\eqref{nmio2} and repeated the calculation in Fig. \ref{figoneside1}. It is found that $\Delta t_{v,ij}$ in both mass ordering scenarios are independent of the masses in the given range. Indeed, for $\Delta t_v$ between ultra-relativistic particles with same energy but different masses, one can replace the velocity in the first term of Eq. \eqref{eqdtvapp} by $v_i=\sqrt{1-m_i^2/E^2}$ and then expand around the small $m_i/E$. One finds that  $\Delta t_{v,ij}$ is equivalent to the time delay found for a single gravitational potential in Ref. \cite{Zatsepin:1968kt}, which is given by
\be
\Delta t_{v,ij}=5.15~\mbox{ms}\cdot \frac{\Delta m_{ij}^2/\mbox{eV}^2}{\left( E/\mbox{10~MeV}\right)^2} \cdot \frac{D_{os}}{\mbox{10~kpc}} \; , \label{tof}
\ee
where $D_{os}$ is the total distance from the SN to the observer.
Clearly, the leading term of the time delay is only sensible to mass square difference but not the absolute value of the masses of the neutrinos.

Indeed, the time delay between different mass eigenstates of neutrinos originating form a gravitational potential has been considered in Ref. \cite{Jia:2017oar} using Eq. \eqref{tof} with more realistic SNN spectrum and neutrino-matter interaction cross-sections. The findings there was similar to what was observed in Fig. \ref{figoneside1} that the mass eigenstates might be separated by different time intervals in different mass orderings. However, it was pointed out that in order to have enough statistics, the distance $D_{os}$ cannot be too large even for a gigaton water or liquid scintillator detector. Limited $D_{os}$ in Eq. \eqref{tof} then implies that in order to have enough temporal resolution, only sharp features  with very short time duration ($\leq 0.1$ ms) in the SNN spectrum can be used for the purpose of discriminating the mass orderings.
Comparing to the case without GL however, the time delay in GL case fortunately do have an important advantage that the neutrino flux is significantly  magnified, up to 100 times \cite{Rubin:2017ipu}. This will strongly increase the detection event rates for sources from the same $D_{os}$ and therefore makes the method more practical.

For SNN signal from different paths, the time delay $\Delta t_\mathrm{p}(v_i)$ between same mass eigenstate $|\nu_i\rangle$ with velocity $v_i$ is given by Eq. \eqref{eq:675}. Comparing this with Eq. \eqref{tof} we can find that for $\beta>10^{-3}$ [as], $\Delta t_\mathrm{p}(v_i)$ is much larger than $\Delta t_{v,ij}$. Therefore the two series of neutrino signals will not have any overlap for most of the ranges of $\beta$. If we consider the difference of two time delays, $\Delta t_\mathrm{p}(v_i)$ of the ultra-relativistic $|\nu_i\rangle$ signal and $\Delta t_\mathrm{p}(c)$ of the optical signal, then Eq. \eqref{eqdtpv} should be used. Using $v_i=\sqrt{1-m_i^2/E^2}$, it becomes
\begin{align}
 \Delta t_{\mathrm{p}v}
=&-\frac{\beta \sqrt{ r_\mathrm{f}(r_\mathrm{i}+r_\mathrm{f}) \left[\beta^2 r_\mathrm{f}(r_\mathrm{i}+r_\mathrm{f})+16M r_\mathrm{i}\right]}}{2r_\mathrm{i}}\frac{m^2_i}{2E^2}. \label{eqdtpv2}
\end{align}
Formally, measuring $ \Delta t_{\mathrm{p}v}$ and $E$ allows the determination of the absolute mass of $|\nu_i\rangle$ for given $\beta, ~r_\mathrm{i/f}$ and $M$. This was not possible when using solely the time difference $\Delta t_v$ because the difference in emission times of neutrino and optical signal in SN cannot be determined very precisely (e.g. SN1987A), while using the difference of two time delays can avoid this uncertainty. Practically however, one can find using reasonable $r_\mathrm{i}$ and $r_\mathrm{f}$, and typical $\beta$ and $E$  that this time difference is too small to be experimentally resolved if the neutrino masses are in the range specified by Eq. \eqref{nmno1} to \eqref{nmio2}. For example, in the large $\beta$ and $r_\mathrm{i/f}$ limit, Eq. \eqref{eqdtpv2} becomes after restoring all units
\be
 \Delta t_{\mathrm{p}v}\approx 6.05\times 10^{-12} [\mathrm{s}] \cdot \frac{\beta^2}{1~[\mathrm{as}]^2} \frac{r_\mathrm{f}}{1~[\mathrm{Mpc}]}\frac{(r_\mathrm{i}+r_\mathrm{f})}{r_\mathrm{i}} \frac{(m_i/1~[\mathrm{ev}])^2}{\lsb E/(10~[\mathrm{MeV}])\rsb^2}.
\ee
Therefore even for features in SNN spectrum that is as narrow as the neutrino observatory uncertainty ($\sim$1 [ns]), to resolve the peaks form different mass eigenstate would require an extremely large $r_f$. This in turn requires extremely large detectors to reach high enough statistics. Therefore until such detectors are built, this practically will not put any constraint on the neutrino absolute mass.

\subsection{Time delay in a general mass profile and time delay of GW \label{tdgpgw}}

In the analysis of the GW170817 and GRB 170817A signal \cite{Monitor:2017mdv}, it was deduced from the $+1.74\pm 0.05$ [s] time delay of GRB that the speed of GW is constrained to the range
\be
-7\times 10^{-16}\lesssim 1-\frac{v_\mathrm{GW}}{c} \lesssim 3\times 10^{-15}. \label{eq:gwv}
\ee
The main uncertainty in Eq. \eqref{eq:gwv} comes from the fact that this time delay is not necessarily the difference of the total travel times of GW and GRB, because  their emission time could also be different. To avoid this problem, Refs.  \cite{Fan:2016swi,Wei:2017emo,Yang:2018bdf} proposed to use the difference between time delays of GW images  and time delay of GRB images in GL to accurately determine the GW velocity.
However, the time delay used in these works  (Eq. (15) of Ref. \cite{Fan:2016swi}, Eq. (1) of Ref. \cite{Wei:2017emo} and Eq. (2) of Ref. \cite{Yang:2018bdf})
was derived for a singular isothermal sphere profile  from the the time delay between lensed and unlensed rays of light \cite{Biesiada:2007rk}
\be
\Delta t_g = \frac{D_{ol}D_{os}}{ D_{ls}}\lsb \frac{1}{2}( \vec{\theta} -\vec{\beta})^2 - \psi( \vec{\theta}) \rsb, \label{eq:tgori}
\ee
but not the time delay of signal of arbitrary velocity. Therefore it requires a revision.
Note in Eq. \eqref{eq:tgori}, redshift $z$ was set to zero because we are in a Schwarzschild spacetime,  $\psi(\vec{\theta})$ is the effective lensing potential at angle $\vec{\theta}$ and $D_{ol}$ and $D_{ls}$ are the distance from observer to lens and lens to source respectively.
In this work we will update this equation, and show that the first (geometric) term in the bracket of Eq. \eqref{eq:tgori} receives a factor of $1/v$, and the second (potential) term gets a factor of $(3 v^2 - 1)/(2 v^3)$ where $v$ is the speed of the lensed signal. In other words, the time delay formula becomes
\be \label{tdang}
\Delta t_g= \frac{D_{ol}D_{os}}{ D_{ls}}\lsb \frac{1}{2v}( \vec{\theta} -\vec{\beta})^2 - \frac{3 v^2 - 1}{2 v^3}\psi( \vec{\theta}) \rsb.
\ee

In order to illustrate this, we only need to consider the time delay caused by a point mass, i.e.,  a Schwarzschild spacetime.
The time delay formula for an light ray in this case was known to be \cite{Keeton:2005jd}
\begin{align} \label{delt}
\Delta t =  \frac{D_{ol} D_{os}}{D_{ls}} \frac{\lb \theta - \beta \rb^2}{2} + 2 M\ln \lb \frac{ 4 D_{ls}}{D_{ol}\theta^2} \rb.
\end{align}
Now for timelike particles, their motion in Schwarzschild metric satisfy the following normalization condition
\begin{align}
1=& \lb 1-\frac{2 M}{r}\rb \lb \frac{\dd t}{\dd \lambda} \rb^2 - \lb 1-\frac{2 M}{r}\rb^{-1} \lb \frac{\dd r}{\dd \lambda} \rb^2 + r^2 \lb \frac{\dd \theta}{\dd \lambda} \rb^2 + r^2 \sin^2\theta \lb \frac{\dd \phi}{\dd \lambda} \rb^2 \nonumber\\
\approx & \lb 1-\frac{2 M}{r}\rb \lb \frac{\dd t}{\dd \lambda} \rb^2 - \lb 1-\frac{2 M}{r}\rb^{-1} \lb \frac{\dd l}{\dd \lambda} \rb^2 , \label{metmass}
\end{align}
where $\lambda$ is the proper time and $l$ is the length parameter along the path $\dd l^2 =\dd r^2+ r^2(\dd \theta^2+\sin^2\theta\dd\phi^2)$ and the approximation is valid because $r\gg M$.
Using the definition of energy $E$ in Eq. \eqref{gdefl} to replace 1 on the left hand side of Eq. \eqref{metmass}, it can be rewritten as\be
\lb 1-\frac{2 M}{r}\rb \lb \frac{\dd t}{\dd \lambda} \rb^2 - \lb 1-\frac{2 M}{r}\rb^{-1} \lb \frac{\dd l}{\dd \lambda} \rb^2 = \frac{1}{E^2}\lb 1 - \frac{2M}{r}\rb^2 \lb \frac{\dd t}{\dd \lambda} \rb^2.
\ee
Introducing Newtonian potential $U(r)=-\frac{M}{r}$ and rearranging the equation, the travel time can be expressed as
\be
t = \int \frac{\dd l}{\lb 1 + 2 U(r)\rb \sqrt{1 -  \lb 1 + 2U(r)\rb/E^2}},
\ee
Since $U(r)$ is small, expanding this equation to the first order of $U(r)$ yields
\bea
 t &=&  \frac{1 }{ \sqrt{1 - 1/E^2 } } \int \dd l - \frac{2 - 3 /E^2 }{ \lb 1 - 1/E^2 \rb^{3/2} } \int U(r)\dd l,\label{tpappre}\\
 &=& \frac{1}{v} \int \dd l - \frac{3 v^2 -1}{ v^3}\int U(r)\dd l,  \label{tpap}
\eea
where $E$ was replaced by $v$ using Eq. \eqref{lbvrel}.

To carry out the integral in Eq. \eqref{tpap}, we do a change of variables from length parameter $l$ to $x$ in Fig. \ref{flensp} by using the geometric relation
\be
r^2 = x^2 + \lb \xi - \frac{\xi '}{D_{ls}} x \rb^2,\ l^2 = x^2 + \lb \frac{\xi '}{D_{ls}} x \rb^2,
\ee
so that the time from the source to lens plane becomes
\begin{align}
t_{ls}=&   \frac{1 }{v }  \int_0^{D_{ls}} \sqrt{1+\frac{\xi ^{\prime 2}}{D_{ls}^2}} \dd x\nonumber\\
&
 + 2M \frac{3 v^2 -1}{2 v^3}  \int_0^{D_{ls}} \sqrt{1+\frac{\xi ^{\prime 2}}{D_{ls}^2}} \lsb  x^2 + \lb \xi - \frac{\xi '}{D_{ls}} x \rb^2\rsb^{-1/2} \dd x.
\end{align}
Because $\xi,~\xi '\ll D_{ls},~D_{ol}$, this integral can be carried out to the leading orders of $\xi/D_{ls}$ or $\xi'/D_{ls}$ to find
 \begin{align} \label{tsl}
t_{ls}  \approx &  \frac{1}{v} \lb D_{ls} + \frac{\xi ^{\prime 2}}{2 D_{ls}} \rb + 2M \frac{3 v^2 -1}{2 v^3} \ln \lb \frac{ 2 D_{ls} }{\xi} \rb.
\end{align}
Similarly, the time from the lens plane to the observer is found to be
\begin{align} \label{tlo}
t_{ol}  \approx &  \frac{1}{v} \lb D_{ol} + \frac{\xi ^2}{2 D_{ol}} \rb + 2M \frac{3 v^2 -1}{2 v^3} \ln \lb \frac{ 2 D_{ol} }{\xi} \rb
\end{align}
and the total travel time becomes
\be \label{tpapf}
t_\mathrm{tot} =  \frac{1}{v} \lb D_{ls} + D_{ol} + \frac{\xi ^{\prime 2}}{2 D_{ls}} + \frac{\xi^2}{2 D_{ol}} \rb + 2 M \frac{3 v^2 -1}{2 v^3} \ln \lb \frac{ 4 D_{ls} D_{ol}}{\xi^2} \rb.
\ee
The time delay between the travel time for the lensed ray and unlensed ray then is
\be \label{tdgw1}
\Delta t_{\mathrm{g}} =  t_\mathrm{tot} - \frac{D_{os}}{\cos \beta}.\ee
Since we are in the weak field limit, the angle $\beta$ and $\theta$ are small. From Fig. \ref{flensp} we have
\be
\theta \approx \tan \theta = \frac{\xi}{D_{ol}},\ \beta \approx \tan \beta = \frac{\xi - \xi'}{D_{os}},
\ee
which lead to a solution of $\xi$ and $\xi'$ in terms of other observables
\be \label{xiterm}
\xi = D_{ol} \theta,\ \xi' = D_{ol} \theta - D_{os} \beta.
\ee
Substituting Eqs. \eqref{xiterm} into Eq. \eqref{tdgw1}, one finally finds
\be
\Delta t_g = \frac{D_{ol} D_{os}}{D_{ls}} \frac{\lb \theta - \beta \rb^2}{2v} + 2M \frac{3 v^2 -1}{2v^3}  \ln \lb \frac{ 4 D_{ls}}{D_{ol}\theta^2}  \rb .
\ee
Note that if we expand $\Delta t_g$ for relativistic velocity, this result is in accordance with Eq. (7) of Ref. \cite{Glicenstein:2017lrm} (although it seems a factor 2 difference occur at the $(1-v)^1$ order).

Comparing to formula \eqref{delt}, one can immediately read off the extra factors due to velocity in the geometric term and the potential term. Since the time delay for general potential is a convolution of point mass potential, then it is apparent that for a general potential $\psi(\vec{\theta})$, the time delay of an relativistic particle with velocity $v$ becomes
\be
\Delta t_g = \frac{D_{ol}D_{os}}{D_{ls}}\lsb \frac{1}{2v}( \vec{\theta} -\vec{\beta})^2 - \frac{3 v^2 - 1}{2 v^3}\psi( \vec{\theta}) \rsb \label{tdangrep}
\ee
which is exactly the desired Eq. \eqref{tdang}.
Note that if the relativistic limit is taken, then the first term agrees with Eq. (19) of Ref. \cite{Baker:2016reh} at the $\calco(1-v)^1$ order. The potential term there however has the same dependance on velocity as the first term in the bracket at order $\calco(1-v)^1$, which is at odds with our result and Ref. \cite{Glicenstein:2017lrm}.

Eq. \eqref{tdangrep} is applicable to general mass profiles to study the difference between two time delays of different signals. However, here we will only not further pursue along this direction but concentrate on the simple point mass result, Eq. \eqref{eqdtpv}. For GW and GRB signal, their time delay difference becomes
\begin{align}
\Delta t_{\mathrm{p}v,\mathrm{GW}}
=&\frac{\beta \sqrt{ r_\mathrm{f}(r_\mathrm{i}+r_\mathrm{f}) \left[\beta^2 r_\mathrm{f}(r_\mathrm{i}+r_\mathrm{f})+16M r_\mathrm{i}\right]}}{2r_\mathrm{i}}( 1-v_\mathrm{GW}). \label{eqdtpvgw}
\end{align}
Clearly, this difference is linear to $1-v_\mathrm{GW}$. Its dependance on $\beta$ and the distance is the same as in Eq. \eqref{eqdtpv} and similar to Eq. \eqref{dtbf} .

\begin{center}
\begin{figure}[htp!]
\includegraphics[width=0.49\textwidth]{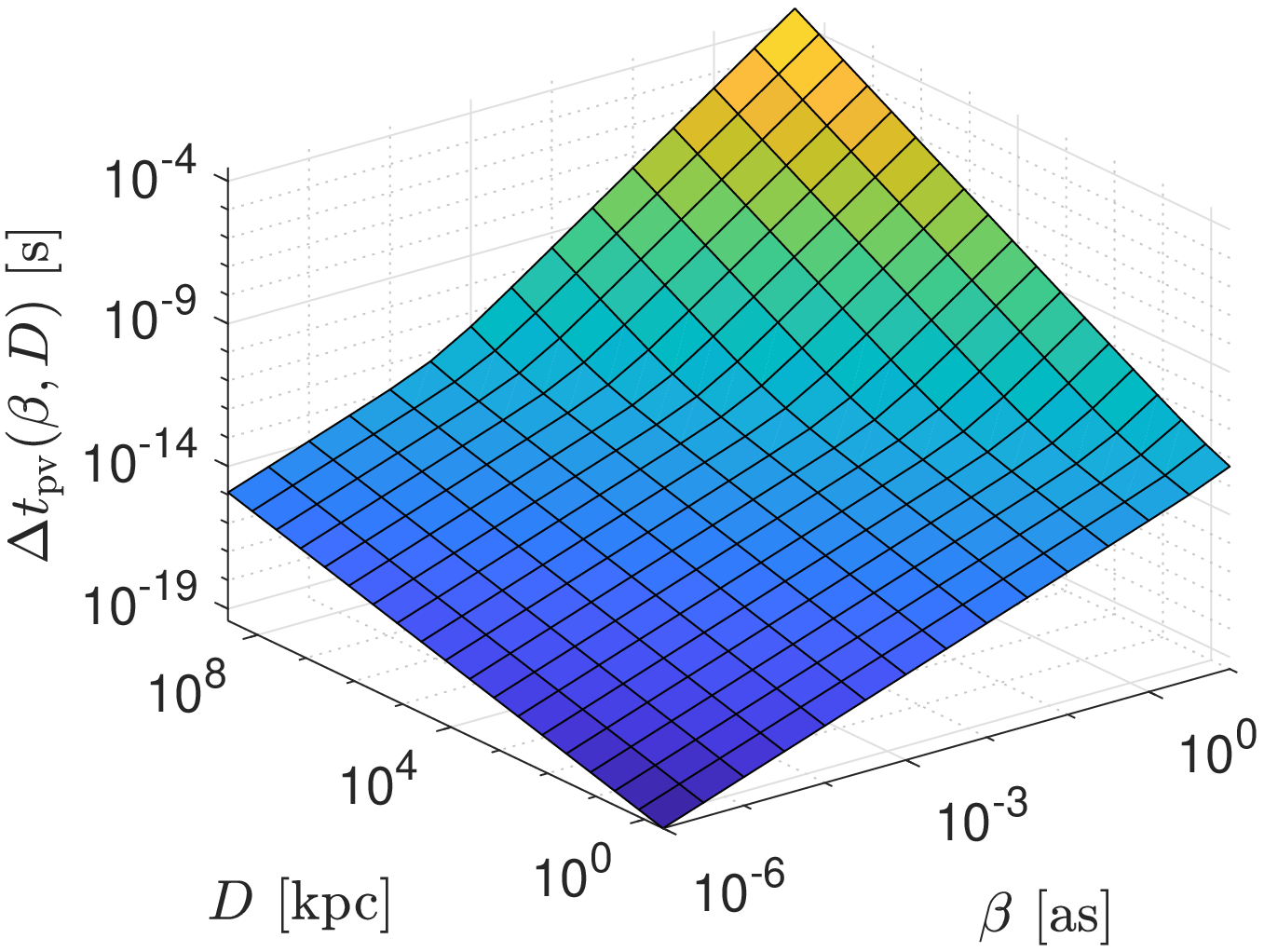}
\includegraphics[width=0.49\textwidth]{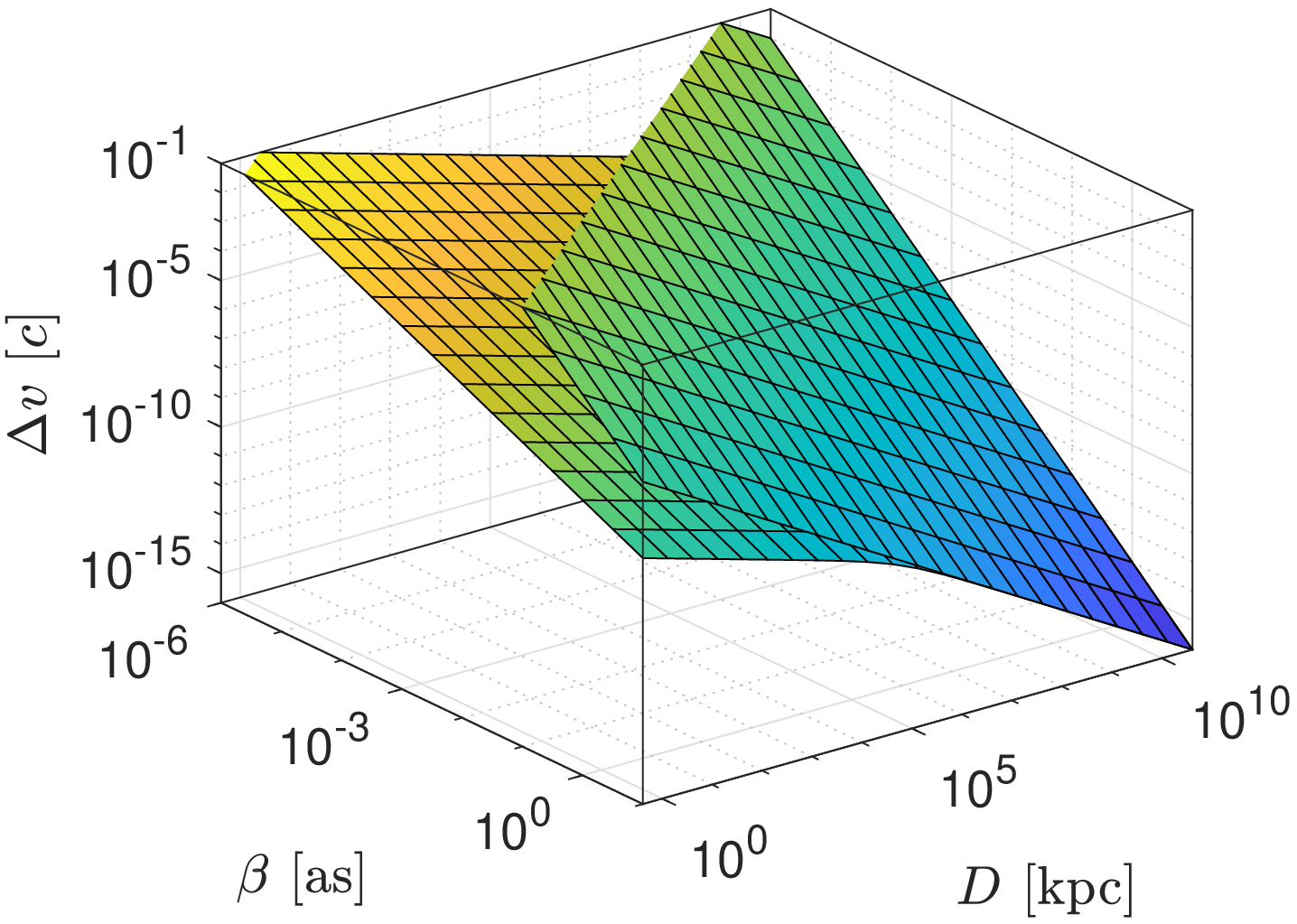}\\
(a)\hspace{6cm}(b)
\caption{\label{figdifftdgw} (a) Difference in time delays of GRB and GW signals as a function of $D$ and $\beta$ for $v_\mathrm{GW}=(1-3\times 10^{-15})c$. (b) The constraint on GW velocity $v_\mathrm{GW}$ as a function of $D$ and $\beta$ assuming that the time delay difference $\Delta t_{\mathrm{p}v,\mathrm{GW}}=1$ [ms] is measured. }
\end{figure}
\end{center}

Since previously $v_\mathrm{GW}$ has been constrained to the range of Eq. \eqref{eq:gwv},
using the maximum deviation of $v_\mathrm{GW}$ from $c$ that is still allowed, i.e., $1-v_\mathrm{GW}/c=3\times 10^{-15}$, we can estimate the difference $\Delta t_{\mathrm{p}v,\mathrm{GW}}$. In Fig. \ref{figdifftdgw} we plot $\Delta t_{\mathrm{p}v,\mathrm{GW}}$ as a function of $\beta$ and $D$ for velocity $\Delta v/c=3\times 10^{-15}$.
It is seen that the parameters considered take their maximal values that are considered, i.e., $D_{ol}=D_{ls}=2\times 10^4$ [Mpc] and $\beta=10$ [as], the difference in GW and GRB time delay also reach maximum, $\Delta t_{\mathrm{p}v,\mathrm{GW}}=1.45\times 10^{-5}$ [s]. This distance corresponds to $z=4.3$ and is already 10 times the maximal distance of currently detected GW candidate \cite{gwevent1} and slightly larger than the largest distance of detected strong optical GLs \cite{opgllist}. Similarly, the value of $\beta$ we used here is also larger than known optical GL values \cite{opgllist}. On the other hand, it is also known that current uncertainties in the measurement of GW event time is 0.002 [s] and that of GRB event time is 0.05 [s] \cite{Monitor:2017mdv}. These numbers are 2 and 3 orders larger than the above difference of $1.45\times 10^{-5}$ [s] at largest $D$ and $\beta$. Therefore, in order to further constraint GW velocity, either these uncertainties have to be improved by 2 and 3 orders respectively for GW and GRB measurements, or some very rare gravitationally lensed binary neutron merger event from a distance larger by a factor of 3 orders, i.e., $\sim 10^7$ [Mpc], has to be detected.

To evaluate the best constraining power to velocity difference $\Delta v\equiv 1-v_\mathrm{GW}/c$ by a given measured $\Delta t_{\mathrm{p}v,\mathrm{GW}}$, in Fig. \ref{figdifftdgw} (b) we plot $\Delta v$ as a function of $\beta$ and $D$ assuming that a 1 [ms] time delay difference was measured by the detectors. The upper surface is for $M=4.12\times 10^{11}M_\odot$ and the lower  surface is for $M=4.12\times 10^{-1}M_\odot$.  It is seen that the larger the $D$ and $\beta$, the smaller the $1-v_\mathrm{GW}/c$ can be constrained. Moreover, at large $D$ and $\beta$, the two surface due to different $M$ converge.  Indeed, taking the large $D$ and $\beta$ limit, one can obtain the velocity difference $1-v_\mathrm{GW}/c$ from Eq. \eqref{eqdtpvgw} as
\be
1-\frac{v_\mathrm{GW}}{c}\approx 4.13\times 10^{-7}\cdot \frac{\Delta t_{\mathrm{p}v,\mathrm{GW}}}{1~[\mathrm{ms}]}
\frac{1~[\mathrm{as}]^2}{\beta^2} \frac{1~[\mathrm{Mpc}]}{r_\mathrm{f}}\frac{r_\mathrm{i}}{(r_\mathrm{i}+r_\mathrm{f})} .
\ee
This shows clearly that constraining power of lens and source at different distance and angular positions.

\section{\label{secdis} Conclusion and Discussions}

The time delay of timelike particles in GL is important for astrophysical applications such as constraining GW velocity and neutrino mass/mass orderings. In this work, we studied the time delay of signals with arbitrary velocity in GL of the Schwarzschild spacetime. Exact formula of the total time $t_\mathrm{if}$  is obtained in  Eq. \eqref{tDS} as elliptical function and the approximation result of $t_{if}$ is found in Eq. \eqref{ttbigyzv} under the weak field limit. Both the exact and approximate $t_\mathrm{if}$ are functions of the gravitational center mass, the source and observer distances, the particle  velocity and minimal radius $r_0$, the last of which is linked to the angular position of the source using lens equation.

The time delay $\Delta t_v$ for signals of different velocities but on the same side of the lens and various limits of $\Delta t_v$ are obtained in Sec. \eqref{subsectdv}. The time delay $\Delta t_\mathrm{p}$ for signals on opposite sides of the lens and its approximations are obtained in Sec. \eqref{sectdtwop}. The dependance of $\Delta t_v$ and $\Delta t_\mathrm{p}$ on various parameters including $\beta$, $v$ and $r_\mathrm{i/f}$ are discussed carefully.

These time delays are applied to the time delay of SNNs and GW. It is shown that the time delay $\Delta t_v$ between relativistic neutrino mass eigenstates $|\nu_i\rangle$ and $|\nu_j\rangle$ is proportional to $D_{os}\Delta m^2_{ij}/E$. Therefore the three mass eigenstates will yield a sequence of neutrino signals that is different for normal and inverted orderings. This implies a possibility to discriminate these orderings, although a very large detector is required to have enough statistics.

For GW application, we first updated the formula of time delay in GL of a general mass profile for signal with arbitrary velocity. Then in Schwarzschild spacetime, it was shown that for distance as large as $2\times 10^4$ [Mpc] and source angle of 1 [as], the difference in time delays of GW signal with velocity $v_\mathrm{GW}=(1-3\times 10^{-15})c$ and GRB signal can only reach $1.45\times 10^{-5}$ [s]. To utilize this difference to further constraint GW velocity, the measurement uncertainty of GW and GRB has to be improved or very high redshift GW/GRB event has to be observed.

It is instructive to comment on the future extensions of the current work. The first is to apply the time delay formula for general mass profile to more specific lens mass distributions and study the corresponding implications on GW properties. The perturbative methods used in the work can also be extended to other spacetimes, preferably those with spherical/axial symmetries, such as Kerr spacetime. It would be interesting to know how the spin angular momentum of the spacetime affects the time delay of massive particles from different images in GL. We are currently working along these directions. 

\begin{acknowledgments}
The authors appreciate the discussion with Dr. Yungui Gong and Xilong Fan. This work is supported by the National Nature Science Foundation of China No. 11504276.
\end{acknowledgments}

\appendix

\section{Definitions of the elliptic functions} \label{appda}

And the special functions appear in the equations are elliptic integral. Then we introduce the definitions of the elliptic integrals. The elliptic integral of the first kind is
\begin{align}
F(\phi|m)=\int_0^{\phi}\lbr 1-m\sin^2\left(\theta\right)\rbr^{-1/2}\dd\theta,\ (-\pi/2<\phi<\pi/2).
\end{align}
The elliptic integral of the second kind is
\begin{align}
E(\phi|m)=\int_0^{\phi}\lbr 1-m\sin^2\left(\theta\right)\rbr^{1/2}\dd\theta,\ (-\pi/2<\phi<\pi/2).
\end{align}
The incomplete elliptic integral is
\begin{align}
\Pi(n;\phi|m)=\int_0^{\phi}\lbr 1-n\sin^2(\theta)\rbr^{-1}\lbrack1-m \sin^2(\theta)\rbrack^{-1/2}\dd \theta.
\end{align}

\end{document}